\documentclass[fleqn,usenatbib]{mnras}

\usepackage{newtxtext,newtxmath}

\usepackage[T1]{fontenc}
\usepackage{ae,aecompl}
\usepackage{soul}

\usepackage{graphicx}	
\usepackage{amsmath}	
\usepackage{amssymb}	
\usepackage[usenames]{xcolor}
\usepackage{natbib}

\def\equationautorefname~#1\null{equation~(#1)}
\newcommand{\autorefp}[1]{%
  \begingroup%
  \def\equationautorefname~##1\null{equation~##1\null}%
  \autoref{#1}%
  \endgroup%
}

\DeclareMathAlphabet{\mathpzc}{OT1}{pzc}{m}{it}\definecolor{purple}{RGB}{160,32,240}
\newcommand{\revv}[1]{\textcolor{black}{#1}}
\newcommand{\rev}[1]{\textcolor{black}{#1}}

\newcommand{\Mh}{M_{\mathrm{h}}}
\newcommand{\Msun}{\,M_{\odot}}
\newcommand{\Msunh}{\,h^{-1}\,M_{\odot}}
\newcommand{\Mgc}{M_{\mathrm{GC}}}
\newcommand{\Mgcobs}{M_{\mathrm{GC,obs}}}
\newcommand{\tdep}{t_{\mathrm{dep}}}
\newcommand{\tform}{t_{\mathrm{form}}}

\newcommand{\Mstar}{M_{\star}}
\newcommand{\Miso}{M_{\mathrm{tr}}}
\newcommand{\ttid}{t_{\mathrm{tid}}}
\newcommand{\tiso}{t_{\mathrm{iso}}}
\newcommand{\fstar}{f_{\star}}
\newcommand{\SFR}{\mathrm{SFR}}
\newcommand{\sSFR}{\mathrm{sSFR}}
\newcommand{\feh}{\mathrm{[Fe/H]}}

\title{Formation of Globular Cluster Systems: From Dwarf Galaxies to Giants}

\author[Choksi]{
Nick Choksi$^{1,2 \href{https://orcid.org/0000-0003-0690-1056}{\includegraphics[scale=0.4]{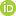}}}$\thanks{E-mail: nchoksi@berkeley.edu},
Oleg Y. Gnedin$^{2 \href{https://orcid.org/0000-0001-9852-9954}{\includegraphics[scale=0.4]{orcid.png}}}$ and
Hui Li$^{2,3 \href{https://orcid.org/0000-0002-1253-2763}{\includegraphics[scale=0.4]{orcid.png}}}$ 
\\
$^{1}$Department of Astronomy, University of California at Berkeley, Berkeley, CA, 94720, USA\\
$^{2}$Department of Astronomy, University of Michigan, Ann Arbor, MI, 48109, USA\\
$^{3}$Department of Physics, Kavli Institute for Astrophysics and Space Research, MIT, Cambridge, MA 02139, USA
}

\date{Released \today}

\pubyear{2018}

\begin{document}
\label{firstpage}
\pagerange{\pageref{firstpage}--\pageref{lastpage}}
\maketitle

\begin{abstract}
Globular cluster (GC) systems around galaxies of a vast mass range show remarkably simple scaling relations. The combined mass of all GCs is a constant fraction of the total galaxy mass and the mean metallicity and metallicity dispersion of the GC system scale up weakly with galaxy mass. The metallicity of massive, metal-poor (``blue") clusters increases with cluster mass, while that of metal-rich (``red") clusters does not. A significant age-metallicity relation emerges from analysis of resolved stellar populations in Galactic GCs and unresolved populations in nearby galaxies. Remarkably, all these trends can be explained by a simple merger-based model developed in previous work and updated here using recent observations of galaxy scaling relations at high redshift. We show that the increasing dispersion of GC metallicity distributions with galaxy mass is a robust prediction of the model. It arises from more massive galaxies having more mergers that combine satellite GC systems. The average metallicity also increases by 0.6~dex over 3~dex in halo mass. The models show a non-linear trend between the GC system mass and host galaxy mass which is consistent with the data. The model does not consider GC self-enrichment, yet predicts a correlation between cluster mass and metallicity for massive blue clusters. The age-metallicity relation is another robust prediction of the model. Half of all clusters are predicted to form within the redshift range $5<z<2.3$, corresponding to ages of $10.8-12.5$~Gyr, in halos of masses $10^{11}-10^{12.5}\Msun$.
\end{abstract}

\begin{keywords}
galaxies: formation --- galaxies: star clusters: general --- globular clusters: general
\end{keywords}


\section{Introduction}
\label{sec:Intro}

Globular cluster (GC) systems in massive galaxies offer unique probes of active star formation at high redshift. Multi-wavelength observations are now available for over a hundred galaxies out to distances of 200~Mpc \citep[e.g.,][]{peng_etal06, harris_etal16_bcg2,harris_etal17_bcg3}. Unlike the integrated light of the whole galaxy, individual clusters provide snapshots of simple stellar populations that reveal the physical conditions at the time of their formation. Especially important is the distribution of metallicities, which is usually inferred from the observed colours. The distribution of cluster masses, inferred from observed luminosities, is also significant, though it is affected by continuous cluster disruption.

GC systems over a vast range of host galaxy masses show bimodality, and generally multi-modality, in their integrated colours. Because all GCs have old stellar populations, this colour bimodality corresponds to a bimodality in the GC metallicity distribution function (MDF): blue GCs are metal-poor, while red GCs are metal-rich. In the Milky Way, only 30\% of the GCs are metal-rich \citep{harris96}. Using a larger sample of the Virgo Cluster Survey (VCS), \citet{peng_etal06} showed that many massive galaxies possess strongly bimodal GC systems, with approximately equal numbers of metal-rich and metal-poor clusters. However, some GC systems are consistent with a unimodal shape, while others appear to be tri-modal and even more complex.

To investigate the origin of the mass and metallicity distributions of GCs in the VCS, \citet[][hereafter LG14]{li_gnedin14} developed a merger-based GC formation model, based on the earlier model of \citet[][hereafter MG10]{muratov_gnedin10}. Using dark matter halo catalogs from the \textit{Millennium-II} simulation and empirically motivated scaling relations to set both the total mass of GCs formed at a given epoch and their metallicities, they showed that bimodality naturally arises from hierarchical mergers. Metal-poor clusters form from many mergers of low-mass halos at high redshift, while metal-rich clusters form from a few high-mass mergers at low redshift.

Recently, studies of GC systems in a diverse population of host galaxies have revealed several new trends. While GC MDFs have historically been modeled as bimodal, the largest systems in brightest cluster galaxies (BCGs) instead show MDFs converging towards unimodality \citep{harris_etal14_bcg1, harris_etal16_bcg2, harris_etal17_bcg3}. The mean metallicity and metallicity dispersion of GC systems both scale with the mass of the host halo (see Fig.~\ref{fig:mean_main}--\ref{fig:std_main}; first noticed by \citealt{peng_etal06} and \citealt{jordan_etal_2007}). Additionally, the combined mass of all globular clusters in a given galaxy shows a near-linear proportionality to the host halo mass over five~orders of magnitude in halo mass \citep{harris_etal_2015}. Also, the massive metal-poor clusters show a trend of increasing metallicity with cluster mass, referred to as the ``blue-tilt'' \citep{strader_etal06, harris_etal06, harris_etal17_bcg3}. Though absolute ages are difficult to measure, observations are finally beginning to reveal a significant age-metallicity correlation \citep{dotter_etal11, georgiev_etal12, leaman_etal13}. The fraction of stellar mass bound in clusters is also consistently a strong function of metallicity \citep[e.g.,][]{harris&harris_2002, beasley_etal_2008, lamers_etal_2017}.

To investigate the origin of these observational trends with host galaxy mass within the merger-based GC formation scenario, we extend the LG14 model to many more halos over a larger range of mass ($10^{11}\Msun \lesssim \Mh \lesssim 10^{14.5}\Msun$). We also update the scaling relations used to set GC properties, motivated by recent observations and hydrodynamical simulations of the cold gas and its metallicity at high redshift. Details of the updates to the model are provided in \autoref{sec:model}. In \autoref{sec:observations} we compile observational data for various GC systems, which we then use to constrain the two free parameters of our model in \autoref{sec:Methodology}. \autoref{sec:Results} reports our predictions for the scaling of GC system properties with host mass. We show that our model reproduces the observed scalings of mean metallicity, metallicity dispersions, and GC system mass. Then we describe the predicted scaling relations for \textit{individual} clusters. Finally, \autoref{sec:discussion} summarizes key results and discusses remaining discrepancies between the model results and observations. We present our conclusions in \autoref{sec:conclusions}.

\section{Updated Model for Globular Cluster Formation} \label{sec:model}

To better follow the assembly of individual dark matter halos, including the merger histories of both central and satellite halos, we switch to using the database of the \textit{Illustris} simulation (\autoref{sec:halo_catalogs}). Stellar masses of the dark matter halos are set by the stellar mass--halo mass relation of \citet{behroozi_etal_2013_main} (\autoref{sec:stellar_mass}). We parameterize both the metallicity of the galaxy and the amount of cold gas available for cluster formation as a function of host stellar mass and redshift, using observed scaling relations (see Sections \ref{sec:metallicity} - \ref{sec:gasfraction}). Cluster formation is triggered in mergers or other accretion events that significantly increase halo mass. Individual clusters are created by Monte Carlo sampling with a standard power-law initial mass function (\autoref{sec:cluster_formation}). Finally, we include cluster mass loss through both stellar evolution and tidal disruption (see \autoref{sec:disruption}). Below we detail all the changes to the model. 

\subsection{Halo Assembly History}
\label{sec:halo_catalogs}

We track the assembly of dark matter halos using the \textit{Illustris-1-Dark} simulation \citep{vogelsberger_etal_2014, nelson_etal_2015}. \textit{Illustris-1-Dark} is a dark matter-only simulation within a $106.5$ Mpc comoving, periodic box. The simulation outputs data at 136 snapshots, from $z=47$ to $z=0$. This provides twice as many outputs as the \textit{Millennium-II} (\textit{MM-II}) simulation used in LG14, allowing for improved tracking of halo merger trees. \textit{Illustris-1-Dark} includes $1820^{3}$ particles at a resolution of $5.3\times 10^{6}\Msunh$. Dark matter (DM) halos were identified using the Subfind algorithm, and their merger trees were constructed using the SubLink algorithm \citep{springel_etal_2001, rodriguez-gomez_etal_2015}. We downloaded the SubLink merger trees of $\approx$200 individual DM halos from the Illustris database and used the fields \textsc{SubhaloMass}, \textsc{FirstProgenitorID}, and \textsc{MainLeafProgenitorID} to connect the halo progenitors through time. To correct for misidentifications by the halo finder, we enforce a monotonic increase in the halo mass ($\Mh$) by skipping simulation outputs at which $\Mh$ decreases. We track all halos with $\Mh > 10^{9}\Msun$, which could host clusters that survive dynamical disruption to the present time (see \autoref{sec:cluster_formation} for details). As in the Illustris Project, we adopt a flat $\Lambda$CDM cosmology with WMAP-9 parameters: $\Omega_{\Lambda} = 0.727$, $h = 0.704$, $n_{s} = 0.963$, $\sigma_{8} = 0.809$ \citep{hinshaw_etal_2013}.

\subsection{Stellar Masses of Dark Matter Halos}
\label{sec:stellar_mass}

LG14 set the stellar mass of dark matter halos at each output of the \textit{MM-II} merger trees via the stellar mass--halo mass (SMHM) relation of \cite{behroozi_etal13b}, with additional scatter. However, because of scatter in the SMHM relation, the resultant time evolution of $\Mstar(t)$ for each halo is not monotonic; halos may experience decreases in $\Mstar$ as large as 0.45 dex between consecutive outputs.

Here, we modify this prescription by setting an initial $\Mstar$ for the first progenitor along each branch of the merger tree via the SMHM relation. The stellar mass at subsequent outputs is then calculated as:
\begin{equation}
  M_{\star,2} = M_{\star,1} + \left(\overline{M}_{\star,2} - \overline{M}_{\star,1}\right) 10^{N(0,\xi)}
  \label{eqn:sm_growth}
\end{equation}
where $\overline\Mstar$ is the median stellar mass at each output. To take into account scatter in the \rev{star formation rate,} we draw $N(0,\xi)$ from a normal-distribution centered on zero with dispersion $\xi$. We apply for $\xi$ the same redshift-dependent parameterization as the scatter in the SMHM relation itself, $\xi(z) = 0.218 + 0.023\, z/(1+z)\,$. This method preserves some memory of the stellar mass at the previous output, such that galaxies that begin with stellar masses deviating from the median trend will in general continue to do so. We have also verified that this method does not violate the SMHM relation at $z=0$.

\subsection{Galaxy and Cluster Metallicities}
\label{sec:metallicity}

In our model, GCs share the metallicity of their host galaxy at the time of formation. The metallicity of the host is in turn set by a stellar mass-metallicity relation (MMR). LG14 adopted a power-law of the form:
\begin{equation}
  \feh =  \log_{10}\left[\left(\frac{\Mstar}{10^{10.5}\Msun}\right)^{\alpha_m} (1+z)^{-\alpha_z}\right],
  \label{eq:mmr}
\end{equation}
with $\alpha_m = 0.4$. However, recent observations suggest the mass-dependence slope was too steep. \citet{kirby_etal13} investigated the MMR in seven dwarf galaxies in the Local Group and find a weaker slope, $\alpha_m \approx 0.3$. \citet{ma_etal_2016} studied the MMR in the FIRE galaxy formation simulations and find the gas metallicity scaling $Z_{\mathrm{gas}} \propto \Mstar^{0.35}$. Motivated by these studies, we adopt a slightly weaker slope for the MMR, $\alpha_m = 0.35$.

The MMR also has a somewhat lower normalization at higher redshift. LG14 adopted a slope for the redshift evolution $\alpha_z = 0.6$. \cite{mannucci_etal_2009} show 0.6 dex evolution in the MMR to redshift $z \approx 4$, yielding a steeper slope $\alpha_z \approx 0.9$. We find that a stronger MMR evolution better reproduces observed GC metallicity distributions, and therefore fiducially adopt $\alpha_z = 0.9$, but also consider the old value as an alternative. 

LG14 adopted an intrinsic metallicity scatter, $\sigma_{\mathrm{met}} = 0.2$ dex. Here, we increase this value to $\sigma_{\mathrm{met}} = 0.3$ dex to better reproduce the observed metallicity dispersion (see \autoref{fig:std_main}). We also update the maximum allowed value of $\feh$ from $+0.2$ to $+0.3$, to account for the most metal-rich stellar populations of giant elliptical galaxies \citep{harris_etal16_bcg2}.

\subsection{Gas Fractions}
\label{sec:gasfraction}

LG14 parametrized the ratio of cold gas to stellar mass in galaxies at high redshift as a double power-law of mass and redshift:
\begin{equation}
  \frac{M_g}{\Mstar} \equiv \eta(\Mstar,z) \propto \Mstar^{-n_m} (1+z)^{n_z}.
\end{equation}
Here, we update this relation. Direct measurements of the full gas content of galaxies at high redshift are still challenging, and ALMA observations are only beginning to sample the molecular $H_2$ gas. A simpler, indirect approach to derive the amount of gas available for star formation is to rewrite the gas fraction as a product of the specific star formation rate, $\sSFR \equiv \SFR/\Mstar$, and the gas depletion time, $\tdep \equiv M_g/\SFR$.

While LG14 adopted a constant $\tdep$ and a single power law for the redshift dependence of $\sSFR(z) \propto (1+z)^{2.8}$ at all $z$, yielding $n_z = 2.8$, recent observations of galaxies at $z \approx 2-3$ show a break in this trend. \citet{lilly_etal_2013} show that the data are consistent with a broken power law: $\sSFR(z) \propto (1+z)^3$ for $z \lesssim 2$ and $\sSFR(z) \propto (1+z)^{1.7}$ for $z \gtrsim 2$.  The gas depletion time for actively star-forming galaxies, on the galactic ``main sequence", also appears to vary with redshift \citep{genzel_etal_2015}. We fiducially adopt $\tdep \propto {(1+z)^{-0.3}}$, but consider in \autoref{subsec:alt_models} an alternate, faster evolution suggested by \citet{tacconi_etal_2017}. We also discuss the implications of possible curvature in the redshift evolution in \autoref{sec:alt_discussion}. For our fiducial model, combining these results gives:
\begin{equation}
  n_z = 1.4 \;\mathrm{for}\; z > 2, \;\mathrm{and}\; n_z = 2.7 \;\mathrm{for}\; z < 2. 
  \nonumber
\end{equation}
In order to connect the gas fractions continuously at $z=2$, we write them as: 
\begin{equation}
  \eta(\Mstar,z) = \eta_9 \, \left(\frac{\Mstar}{10^9\Msun}\right)^{-n_m(\Mstar)} \ \left(\frac{1+z}{3}\right)^{n_z(z)}.
  \label{eqn:fg}
\end{equation}
There is little data on the gas fraction at very high redshift, $z > 3$. Therefore, following LG14, we adopt a fixed upper limit set at $z=3$: $\eta(\Mstar,z>3) = \eta(\Mstar,z=3)$.

Even though the parametrization adopted by LG14 for $\eta(\Mstar,z=0)$ is a good match to the total neutral gas fraction observed in nearby galaxies, it may not be directly relevant for our model. Globular clusters form at higher redshift, $z > 1$, and the model results depend only on the parametrization of the gas fraction at high $z$. The molecular gas fractions of star-forming galaxies at $z \sim 3$, summarized by \citet{tacconi_etal_2017}, correspond to a factor of 5 lower overall normalization than what would follow from the scaling of \autoref{eqn:fg} from $z=0$ with $(1+z)^{n_z}$. Therefore, we revise the normalization factor $\eta_{9}$ accordingly, to be valid at $z > 1$. Our new normalization at $\Mstar = 10^9\Msun$ and $z=2$ is given by: 
\begin{equation}
  \eta_{9} = 0.35\times 3^{n_z(z<2)}.
\end{equation}
Note that it would need to be adjusted for a different scaling slope $n_z(z<2)$, to maintain continuous gas fractions through the break at $z=2$.

For the mass dependence of the molecular gas fraction, \citet{tacconi_etal_2017} fit a single power law with slope $n_m \approx 0.33$. Since the Tacconi sample compilation covers the stellar mass range $10^{9} - 10^{12}\Msun$, we adopt this value of $n_m$ for $\Mstar > 10^{9}\Msun$. The gas content of lower-mass galaxies is probed only at low redshift, where the slope is more shallow, $n_m \approx 0.19$. Since there is no observational revision for these galaxies, we follow LG14 and keep the break in the mass slope at $10^{9}\Msun$. Our resulting combination is
\begin{equation}
  n_m = 0.33 \;\mathrm{for}\; \Mstar > 10^{9}\Msun, \;\mathrm{and}\; n_m = 0.19 \;\mathrm{for}\; \Mstar < 10^{9}\Msun.
  \nonumber
\end{equation}

The resulting parametrization of the gas fraction is roughly consistent with the available data from \citet{tacconi_etal_2017}, but systematically larger by a factor $1.3-2.2$ at all stellar masses $10^{7.5} - 10^{11}\Msun$ and the redshift range $z \approx 2-4$ that are most relevant for globular cluster formation. Some of the remaining discrepancy may be attributed to yet undetected cold gas in high-redshift galaxies. Note that only the functional dependence of $\eta$ on $\Mstar$ and $z$ affects model results.  The overall normalization of the gas fraction ($\eta_{9}$) is degenerate with the parameter $p_2$, which we optimize in \autoref{sec:Methodology}. Our revised gas fractions are a factor of $1.7-8.5$ smaller than those in LG14 across the whole mass and redshift range.  Therefore, we expect $p_2$ to increase by a similar factor, relative to the LG14 best value of $p_2 \approx 2.6$.

Finally, the total baryon fraction cannot exceed the fraction of baryons, $f_{\rm in}$, that can be accreted onto the halo in the presence of the extragalactic UV background: 
\begin{equation}
  \fstar(z) + f_g(z) \leq f_{\rm in}(z),
\end{equation}
where $\fstar \equiv \Mstar/(f_b\Mh)$, $f_g \equiv M_{g}/(f_b\Mh)$, and $f_b$ is the universal baryon fraction. The function $f_{\rm in}(z)$ is computed using equations (2)-(4) in MG10. If the sum $\fstar + f_g$ exceeds $f_{\rm in}$, we set $f_g = f_{\rm in} - \fstar$. This constraint primarily affects small halos ($\Mh \lesssim 10^{10}\Msun$) at $z \gtrsim 9$, in which very few of the surviving clusters form ($\approx$ 3\%; see \autoref{fig:formation1}).

\subsection{Triggers of Cluster Formation}
\label{sec:cluster_formation}

\revv{In our model, we assume that the formation of massive clusters is triggered by periods of rapid accretion onto dark matter halos. Although not explicitly required, such periods are typically caused by major mergers. This simple ansatz, first proposed by \cite{ashman_zepf92}, is motivated by observations of young massive clusters in the nearby universe, which tend to be found in interacting galaxies \citep[e.g.,][]{wilson_etal_2006, portegies_zwart_etal10}. Theoretical studies have found similar results, where mergers induce the high densities and pressures required for massive cluster formation \citep{bournaud_etal08, murray_etal10,kruijssen2015,li_etal17, el-badry_etal_2018}.}

To identify a merger in the \textit{MM-II} halo catalogs, LG14 calculated the \rev{merger ratio, $R_m$, as} the mass of the secondary progenitor, if it existed, to that of the main progenitor. If no secondary progenitor existed \rev{in the catalogs}, they used the differential increase in halo mass \rev{of the main progenitor} between two consecutive outputs, $\Delta \Mh / \Mh$. Cluster formation was triggered if the merger ratio exceeded some threshold value, $p_3$.
\footnote{Even though our current model has only two adjustable parameters, we keep the same notation, $p_2$ and $p_3$, for consistency with previous work.} 

However, the \rev{definition of the merger ratio outlined above} does not take into account the duration of cosmic time between the consecutive simulation outputs. The \textit{Illustris} outputs are more frequent, which would require a larger value of $p_3$ (smaller probability of triggering a formation event per output) to reproduce the same model results. This makes the model parameters dependent on the particular choice of halo merger trees. Also, the time periods between consecutive outputs are typically not constant, which leads to additional temporal dependence.

To fix these issues, we redefine the merger ratio using the logarithmic halo mass accretion rate, $(\Delta M/M)/\Delta t$. For a halo at output time $t_2$ with mass $M_{h,2}$ and its main progenitor at $t_1$ with mass $M_{h,1}$, we write the merger ratio as:
\begin{equation}
  R_m \equiv \frac{M_{h,2} - M_{h,1}}{t_2 - t_1} \frac{1}{M_{h,1}}.
\end{equation}
In our new model, cluster formation is triggered at time $t_2$ if $R_m > p_3$. Thus, we have simplified the model by not searching for a secondary progenitor and instead using only the differential increase of the main progenitor mass.

The new definition of \rev{the merger ratio has} different units than in LG14. \rev{As a rough estimate,} we expect that the new value of $p_3$ will be close to $0.33/\Delta t_{\mathrm{med}}$, where 0.33 is the best value of $p_3$ found in LG14 and \rev{$\Delta t_\mathrm{med, MM-II}$} is the median time between the relevant outputs in the \textit{MM-II} simulation (the adopted dark-matter simulation in LG14) that lead to cluster formation. Cluster formation begins as early as $z \approx 14$ and ends near $z \approx 1$. Over this redshift range, we find $\Delta t_{\mathrm{med}} \sim 0.2$~Gyr, yielding an \rev{estimate} of $p_3 \sim 1$~Gyr$^{-1}$ under our new definition of the merger ratio. \rev{Deviations from this estimate should then mostly due to changes in the adopted scaling relations. After adjusting $p_3$ to match observations (see further discussion in \autoref{sec:Methodology}) we find best values which differ from this estimate by a factor of $\lesssim 2$.}

The episodes when clusters can form are limited in our model by the available outputs of an adopted cosmological simulation. Because of this discreteness, we cannot follow the creation of individual clusters, but rather have to model a small population of clusters formed within some $\Delta t$. This population is characterized by the combined mass of the clusters, $\Mgc$.  We use the same relation as in LG14 between $\Mgc$ and the mass of cold gas available for star formation in the host galaxy, which was originally based on the predictions of galaxy formation simulations of \citet{kravtsov_gnedin05}:
\begin{equation}
  \Mgc = 1.8\times 10^{-4}\, p_2 \, M_g.
  \label{eqn:mgc}
\end{equation}
Here, $p_2$ is the second adjustable model parameter. 

Given the total mass to be formed in clusters, $\Mgc$, we create individual clusters by drawing their masses from a cluster initial mass function (CIMF) of the form:
\begin{equation}
  \frac{dN}{dM} = M_0M^{-2},
  \label{eq:cimf}
\end{equation}
where $M_0$ is a normalization constant. \rev{This power-law shape is consistently observed in nearby spiral and interacting galaxies \citep[e.g.][]{portegies_zwart_etal10}. For simplicity, in this paper we ignore the exponential cutoff of the cluster mass function at very high mass \citep[e.g.,][]{gieles_etal_2006a, larsen2009}, but will present a complete description of its effects in an upcoming paper (N. Choksi \& O. Gnedin 2018, in prep).}  

\rev{Our procedure for sampling the CIMF is based on the ``optimal sampling'' technique described in \cite{schulz_etal_2015}. First, we draw the most massive cluster, of mass $M_{\mathrm{max}}$. The value of $M_{\mathrm{max}}$ is determined from the constraint that there be only one cluster of this mass, i.e.,
\begin{equation}
  1 = \int_{M_{\mathrm{max}}}^{\infty} \frac{dN}{dM} dM.
\end{equation}
Integrating the above gives $M_{\mathrm{max}} = M_0$. In turn, $M_0$ is found from the constraint:
\begin{equation}
  \Mgc = \int_{M_{\mathrm{min}}}^{M_{\mathrm{max}}} M\frac{dN}{dM} = M_{\mathrm{max}}\ln\frac{M_{\mathrm{max}}}{M_{\mathrm{min}}},
\end{equation}
which we solve numerically for $M_0$. We set the minimum cluster mass that can form as $M_{\mathrm{min}} = 10^5\Msun$, because clusters with lower initial mass are expected to be completely disrupted within a few Gyr. Other clusters with $M < M_{\mathrm{max}}$ are drawn using the transformation method (\S7.3 in \citealt{numerical_recipes}). Sampling continues until the mass in generated clusters equals $\Mgc$.}

\rev{Adopting a minimum cluster mass also implies a minimum halo mass capable of forming one cluster. By setting $\Mgc = M_{\mathrm{min}}$ in \autoref{eqn:mgc} and taking $M_g < f_b\Mh$, we find $M_\mathrm{h,min} > 3.3\times 10^{9} p_2^{-1}\Msun$. For the best values of $p_2 \sim 10$ that we find in \autoref{sec:Methodology}, this gives the lower limit $M_\mathrm{h,min} \gtrsim 3\times 10^8 \Msun$. As noted in \autoref{sec:halo_catalogs}, for computational tractability we follow only halos with $\Mh > 10^9\Msun$. We have verified that halos in the mass range $10^8 \Msun \lesssim \Mh \lesssim 10^9 \Msun$ -- in which clusters can in principle form, but that we do not track -- contribute $\lesssim 1\%$ to the total number of clusters. This result is unsurprising, given Eq. \ref{eqn:mgc}: in this range most halos would be able to form only one cluster with $M \sim M_{\mathrm{min}}$ and therefore any cluster the halo forms would be destroyed quickly.}

\rev{The adopted minimum mass of GCs also sets a minimum GC metallicity. For the minimum halo mass calculated above, $M_{\mathrm{h,min}} \sim 3 \times 10^8 \Msun$, the corresponding stellar mass is $\Mstar \sim 6 \times 10^4 \Msun$  at $z \sim 10$  \citep{behroozi_etal_2013_main}, when clusters first begin forming. Using our adopted mass-metallicity relation (\autoref{eq:mmr}) gives the minimum metallicity of a galaxy able to form a GC, $\feh_{\mathrm{min}} \approx -2.9$. This value represents a minimum because higher mass halos will have a higher stellar mass, and therefore a higher metallicity, by the stellar mass-metallicity relation.}

\rev{Field stars require a much smaller gas reservoir than GCs to form, and therefore field stars can form in lower mass and lower metallicity halos. Thus, even though the typical metallicity of GCs is much lower than that of the field stars, the \textit{minimum} metallicity of GCs is predicted to be higher. Observations of the metallicity distribution functions of GCs and halo stars in the Milky Way and other nearby galaxies show qualitatively similar trends \citep[e.g.,][]{lamers_etal_2017}.}

\subsection{Cluster Disruption}
\label{sec:disruption}

After creating clusters, we calculate their stellar and dynamical evolution. We update the dynamical disruption prescription of LG14 with the modified version of \citet{gnedin_etal14}, combining the disruption rates in isolation and in the presence of a strong external tidal field:
\begin{equation}
  \frac{dM}{dt} = -\frac{M}{\mathrm{min}\left(\tiso, \ttid\right)}.
  \label{eqn:dmdt}
\end{equation}

The tidally-limited disruption timescale, $\ttid$, is found by the $N$-body simulations of \citet{gieles_baumgardt08} to be:
\begin{equation}
  \ttid \approx 10\,\mathrm{Gyr}\, \left(\frac{M(t)}{2 \times 10^{5}\Msun}\right)^{2/3} P,
  \label{eqn:ttid}
\end{equation}
where $P$ is a normalized rotation period about the galactic center that accounts for variations in the strength of the local tidal field. Because our model has no spatial information of clusters, we adopt a constant value of the normalization. A choice of $P=1$ corresponds to the median sizes of the Galactic GCs and tidal fields in the solar neighborhood, but we find that models with $P=1$ cannot reproduce observed cluster mass functions in the VCS, and lead to mean cluster masses that are too low by $\approx 0.5$~dex. Furthermore, newly formed GCs are expected to be subjected to strong tidal shocks in the dense interstellar medium (ISM) of merging high-redshift galaxies \citep[e.g.,][]{kruijssen12, kim_etal_2018}. Lower values of $P$ correspond to \textit{stronger} tidal fields and more tidal disruption\rev{, and can also raise the mean cluster mass by completely disrupting very low mass clusters.} Therefore, we adopt a lower value for the normalization, $P=0.5$, and keep it fixed throughout the model calculation. This choice produces reasonable cluster mass functions, as we show in \autoref{sec:gcmf}.

\rev{Evaporation of unbound stars from clusters continues even in weak tidal fields, such as in the halos of galaxies. In this limit, the timescale for evaporation "in isolation", $\tiso$, was calibrated by \citet{gnedin_etal14} as a multiple of the relaxation time evaluated with the typical size-mass relation for Galactic globular clusters:}
\begin{equation}
  \tiso \sim 17\,\mathrm{Gyr}\left(\frac{M(t)}{2 \times 10^{5}\Msun}\right).
	\label{eqn:tiso}
\end{equation}
\rev{Note that this expression does not account for the expansion of the cluster as it relaxes, which significantly increases the length of the relaxation time \citep[e.g.,][]{baumgardt_etal02}.  Therefore the normalization of $\tiso$ in Eq.~\ref{eqn:tiso} is likely too low. However, we have verified that any changes to the normalization, or elimination of disruption in isolation entirely, have little consequence for the final cluster population. This result is unsurprising, for reasons we discuss below.}

Because $\tiso \propto M$, whereas $\ttid \propto M^{2/3}$, there exists a transition mass, $\Miso$, below which $\tiso < \ttid$ always. Equating the two timescales yields:
\begin{equation}
  \Miso = 5.1 \times 10^{3}\Msun \left(\frac{P}{0.5}\right)^3.
\end{equation}
\rev{Thus, disruption in isolation affects only the lowest mass clusters.} Because the minimum mass of clusters created in our model is $10^5\Msun$, no model clusters initially satisfy $\tiso < \ttid$. As a result, we evolve clusters to the transition mass $\Miso$ in the tidally limited regime, and then continue to evolve the cluster in isolation to $z=0$. The transition between the two regimes occurs at a time $t_{\rm tr}$ given by:
\begin{equation}
  t_{\rm tr}(M_0) = \frac{3}{2}\left[1 - \left(\frac{\Miso}{M_0}\right)^{2/3}\right]t_{\mathrm{tid, 0}}, 
\end{equation}
where $t_{\mathrm{tid,0}} \equiv \ttid(M_0)$ is the tidal disruption time for the initial cluster mass $M_0$. Then, integrating \autoref{eqn:dmdt}, we have the mass evolution due to disruption alone: 
\begin{equation}
  M'(t) = \begin{cases} M_0\left[ 1 - \frac{2}{3}\frac{t}{t_{\mathrm{tid,0}}} \right]^{3/2} &\text{for}\ t < t_{\rm tr},\\ 
  \Miso - \left[ \frac{ t - t_{\rm tr}}{17\,\mathrm{Gyr}} \right] 2 \times 10^5\Msun &\text{otherwise.}  
\end{cases}
\end{equation}

We also adopt the time-dependent mass loss rate due to stellar evolution, $\nu_{\rm se}$, derived in \cite{prieto_gnedin08}. Combining both stellar and dynamical evolution, and assuming the dynamical evaporation timescale is much longer, yields a final cluster mass:
\begin{align}
  M(t) = M'(t) \left[ 1 - \int_0^t \nu_{\rm se}(t')dt' \right].
\end{align}

\subsection{Summary of Changes}

To summarize, the main changes in our fiducial model relative to LG14 are:
\begin{enumerate}
\item Switch from \textit{Millennium-II} to \textit{Illustris} halo catalogs.
\item Revised $\Mstar(\Mh,z)$ prescription to ensure consistent stellar mass growth.
\item Slope of the mass dependence of the galaxy MMR reduced from $\alpha_m = 0.4$ to 0.35.
\item Slope of the redshift evolution of the MMR increased from $\alpha_z = 0.6$ to 0.9.
\item Scatter in GC metallicity increased from $\sigma_{\mathrm{met}} = 0.2$ to 0.3 dex.
\item Maximum metallicity increased from $\feh=$ +0.2 to +0.3.
\item Slope of the mass dependence of the cold gas fraction reduced from $n_m = 0.68$ to 0.33 at $\Mstar > 10^{9}\Msun$.
\item Slope of the redshift evolution of the cold gas fraction reduced from $n_z = 2.8$ to 1.4 at $z>2$.
\item Redefinition of the merger ratio, $R_m = (\Delta M/M)/\Delta t$.
\item Revised calculation of dynamical disruption.
\end{enumerate}
The other model ingredients are unchanged and listed in \autoref{sec:cluster_formation}: cluster formation efficiency and cluster initial mass function.

The two alternative models we consider have $n_z(z>2) = 1.1$ and $\alpha_z = 0.6$, respectively (\autoref{tab:models}).

\begin{table*}
\centering
\begin{tabular}{lccccc|ccc}
\hline\\[-2mm]
Model & $p_2$ & $p_3\,(\mathrm{Gyr}^{-1})$ & $n_z(z<2)$ & $n_z(z>2)$ & $\alpha_z$ & $G_{Z}$ & $G_{M}$ & ${\cal M}$ \\[1mm]
\hline\\[-2mm]
Best    & 6.75 & 0.50 & 2.7 & 1.4 & 0.9 & 0.58 & 0.67 & 7.3 \\
Alt Gas & 11.0 & 0.55 & 2.4 & 1.1 & 0.9 & 0.58 & 0.67 & 7.4 \\
Alt MMR & 12.0 & 0.80 & 2.7 & 1.4 & 0.6 & 0.54 & 0.65 & 8.4 \\
\hline
\end{tabular}
\caption{\small Model parameters. ``Alt Gas" has a weaker evolution of the gas fraction, according to \autoref{eqn:fg}. ``Alt MMR" has a stronger redshift evolution of the MMR, according to \autoref{eq:mmr}. Further discussion of the three models is given in \autoref{subsec:alt_models}.}
  \label{tab:models}
\end{table*}

\section{Observational Data}
\label{sec:observations}

We combine globular cluster colors from several datasets for host galaxies with stellar masses ranging from $10^{9}\Msun < \Mstar < 10^{12.5}\Msun$. The VCS provides GC colors for 64 galaxies in this mass range \citep{peng_etal06}. At the highest masses, \citet[][hereafter H14]{harris_etal14_bcg1} give GC colors in 7 BCGs outside 100 Mpc. We also include GC colors for the Milky Way \citep{harris96} and M31 systems \citep{huxor_etal_2014}. Host galaxy stellar masses are calculated using the following color-dependent mass-to-light ratios found in Table A7 of \cite{bell_etal03}:
\begin{align}
\log_{10}\frac{M}{L_K} &= -0.138 + 0.047\times (g-z) ,\\
\log_{10}\frac{M}{L_V} &= \begin{cases} -0.628 + 1.305\times (B-V),\\
 -0.633 + 0.816\times (B-R). \end{cases}
\label{eqn:ml}
\end{align}

Because direct spectroscopic measurements of GC metallicities are difficult outside the Local Group, we convert the VCS GC colors to metallicities using an empirically determined color-metallicity relation. In another departure from LG14, we adopt a non-linear transformation, proposed recently by \cite{harris_etal17_bcg3}:
\begin{equation}
  (g-z) = 1.513 + 0.481 \,\feh + 0.051 \,\feh^2.
  \label{eq:color-met}
\end{equation}

Metallicities of clusters from the HST-BCG survey of H14 were computed using the color-metallicity transformation of \cite{harris_etal06}:
\begin{equation}
  (B-I) = 2.158 + 0.375 \,\feh.
  \label{eq:color-met2}
\end{equation}

We also add data for eight GC systems from \citet[][hereafter H16]{harris_etal16_bcg2}. Instead of the mean and dispersion for the full GC system of those galaxies, Table 4 of H16 gives the parameters of the bimodal split into red and blue clusters: relative fractions, standard deviations of the two modes, and the difference between the mean metallicities. The individual values of the mean metallicities are not given because of possible differences in calibrations. To include these systems in our compilation, we assume the mean metallicity of the blue clusters is located at a fixed $\feh = -1.15$, consistent with results from H14. We then reconstruct the mean and dispersion of metallicities of the entire GC system for each galaxy by summing two Gaussian distributions of the red and blue clusters.

Combining several independent data sets creates some discrepancies. The VCS and BCG surveys use different metallicity calibrations. Additionally, summing the two Gaussians underestimates the true dispersion of the H16 systems. To account for these differences, we adopt the VCS metallicity calibration as our standard and assume the differences in dispersions between any two datasets, for the same galaxy, are linearly related. We then take advantage of the overlap between datasets. M87 is present in the VCS and full H14 data. Additionally, NGC 6166 is present in H16 and in the full dataset of H16, allowing us to transform between the two, as follows:
\begin{align}
  \sigma_{Z,H16} & = k_1 \, \sigma_{Z,VCS} \quad\mathrm{for\ M87}\nonumber \\
  \sigma_{Z,H16} & = k_2 \, \sigma_{Z,H14} \quad\mathrm{for\ NGC\, 6166}
  \label{eqn:z_conversion}
\end{align}
We find $k_1 = 0.805$ and $k_2 = 0.87$, and use these values to scale all metallicities to the VCS calibration.

Finally, we discard extreme metallicities in both the observed and model samples by imposing a universal cut of $-2.3 < \feh < 0.3$. This cut affects $\lesssim 5\%$ of model and observed clusters. We also discard entirely any model or observed galaxies which have a small sample size of clusters, requiring $N_{\rm GC} > 5$.

\section{Methodology of model selection}
\label{sec:Methodology}

While LG14 applied their model to 20 halos in the mass range similar to that of VCS galaxies, we follow $\approx$200 individual \textit{Illustris} halos, with \rev{$z=0$} masses ranging from $10^{11}\Msun$ to $10^{14.5}\Msun$, to better analyze how GC properties scale with host mass. 

\subsection{Merit Function}
\label{subsec:merit}

Our model has only two adjustable parameters: $p_2$ and $p_3$ (the numbering is left consistent with LG14). To help identify a best set of parameters, we define the ``merit function" as:
\begin{equation}
  {\cal M} \equiv
  \frac{1}{N_h} \sum_h \left(\frac{\Mgc(z=0)}{\Mgcobs(\Mh)} - 1 \right)^2
  + \frac{1}{N_h}\sum_{h}  \left( \frac{0.58}{\sigma_{Z, h} } \right)^2
  + \frac{1}{G_{M}} + \frac{2}{G_{Z}}
  \label{eqn:merit}
\end{equation}
where the sum is taken over all halos $h$, and look to minimize ${\cal M}$.

The first term gives the reduced $\chi^2$ of the total GC system mass at $z=0$ vs. the observed relation with host halo mass, obtained recently using weak-lensing mass estimates \citep[e.g.,][]{hudson_etal_2014, harris_etal_2015}:
\begin{equation}
  \Mgcobs \approx 3.4 \times 10^{-5}\, \Mh.
  \label{eq:mgcobs}
\end{equation}
In addition to the errors of the mass estimates, the linear relation has an intrinsic scatter of 0.2~dex. This term of the merit function forces the model to match the overall normalization of $\Mgc$.

The second term weights the dispersion (standard deviation) of the metallicity distribution by the observed value of 0.58~dex. In \autoref{fig:std_main}, we show that the model tends to underestimate the metallicity dispersion within a given galaxy, so that the average dispersion never exceeds the observed value. This term is designed to bring the model dispersion closer to that of the observations.

The final two terms weight the ``goodness" of the full metallicity and cluster mass distributions, which we refer to as $G_Z$ and $G_M$ respectively. As in LG14, we compute the goodness by first constructing pairs of observed galaxies and corresponding model halos. To do so, we convert observed galaxy stellar massses to median halo masses using the SMHM relation. Then, we select all matching \textit{Illustris} halos with a mass within $\pm 0.3$ dex to account for scatter in the stellar mass-halo mass and mass-to-light ratios. For each observed-model pair, we calculate the Kolmogorov-Smirnov (KS) test probability, $p_{KS}$, of the observed and model GC metallicity and mass samples being drawn from the same distribution. The metallicity comparison is made for all galaxies, whereas the mass comparison is against only the VCS galaxies because of data availability.

Because observations of GC systems are incomplete at the low-mass tail, when comparing cluster mass functions via the KS test we only consider clusters with masses above a certain threshold mass. The completeness limit of the VCS is at approximately two magnitudes below the turnover in the GC luminosity function \citep{jordan_etal_2007} \revv{and it corresponds} to a mass of $\log_{10}M/\Msun \approx 4.5$. We therefore exclude all clusters below this mass when making our comparison to observed systems. 

We repeat this process for each observed galaxy, defining $G_{Z}$ and $G_M$ as the fraction of pairs with an acceptable value $p_{KS} \geq 1\%$ for the metallicity and mass samples respectively. Since matching the full metallicity distribution is more important than matching just the dispersion or normalization, we use the factor of 2 to place more emphasis on the final term in \autoref{eqn:merit}. 

\begin{figure}
\includegraphics[width=\columnwidth]{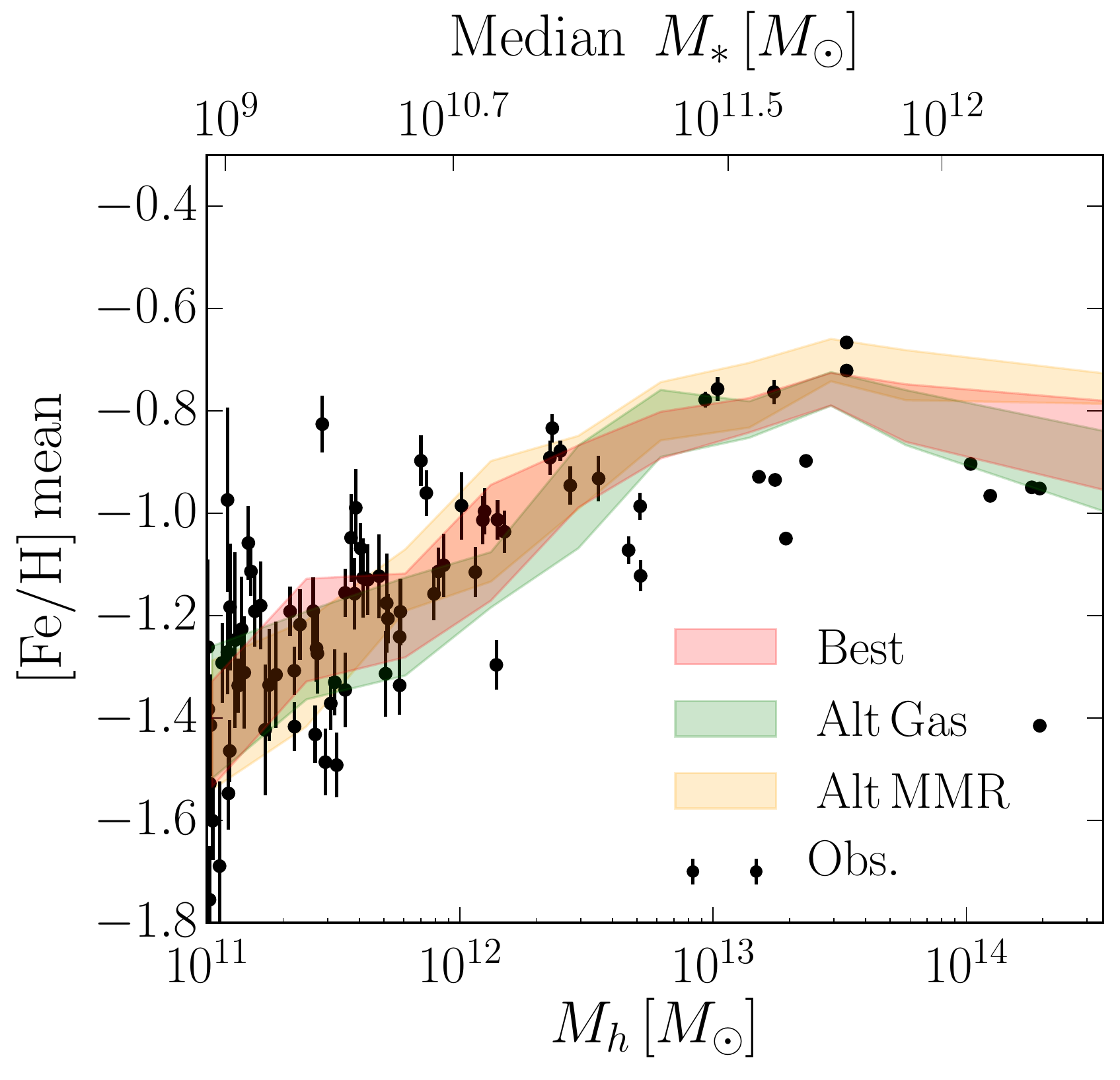}
\vspace{-5mm}
\caption{Mean $\feh$ of GC systems as a function of the galaxy halo mass (\textit{lower scale}) or median stellar mass per halo mass bin (\textit{upper scale}). Shaded regions give the interquartile range (25th to 75th percentiles) of the three models, summarized in \autoref{tab:models}. Red shading denotes the best model, green is the alternative gas evolution, and yellow is the alternative mass-metallicity relation. Black symbols represent data for observed systems, with errorbars showing the standard error of the mean.}
  \label{fig:mean_main}
\end{figure}

\begin{figure}
\includegraphics[width=\columnwidth]{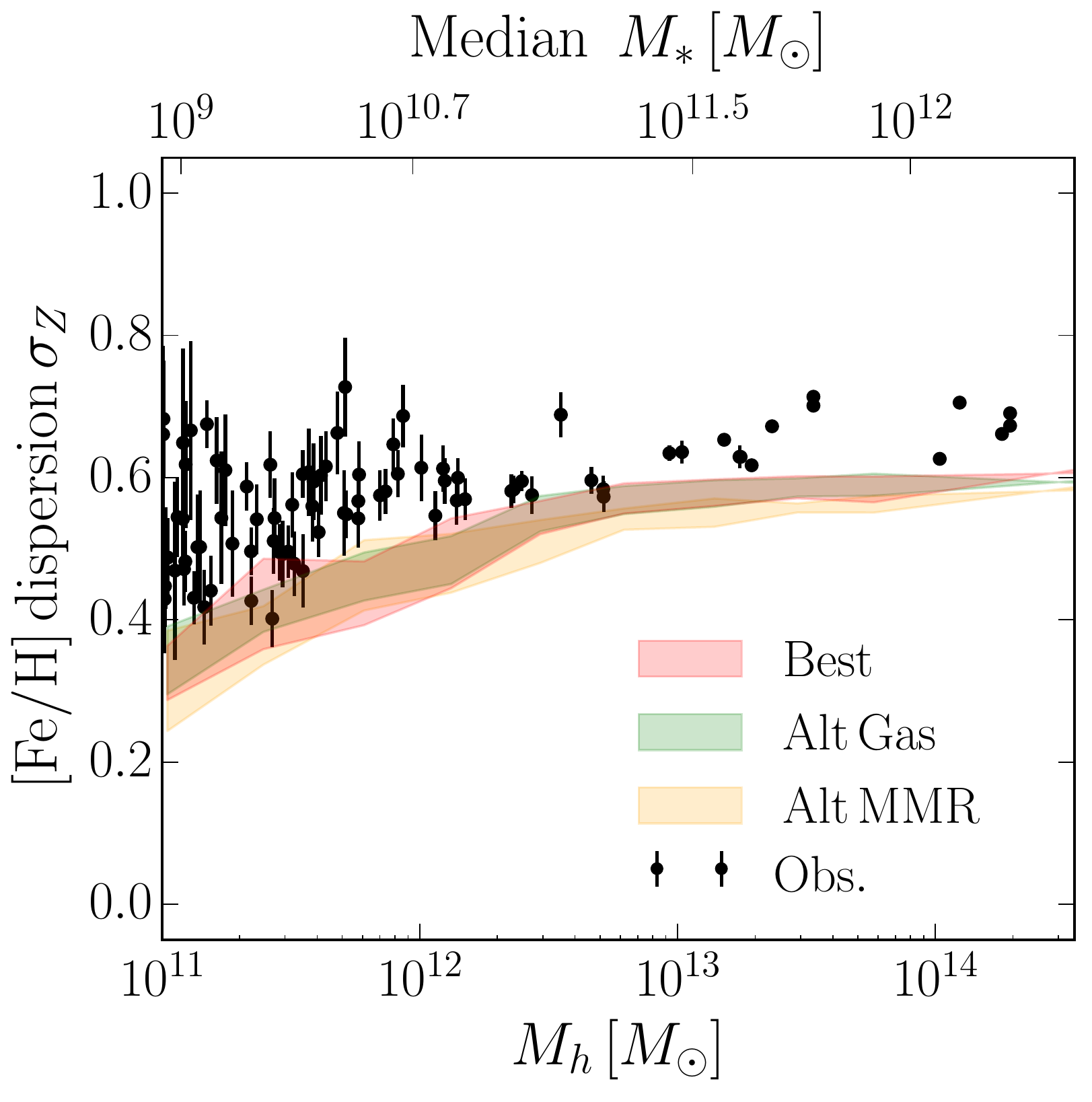}
\vspace{-5mm}
\caption{Standard deviation of $\feh$ of all GCs in a galaxy as a function of the galaxy halo mass (\textit{lower scale}) or median stellar mass per halo mass bin (\textit{upper scale}). Shaded regions give the interquartile range of the three models. Legend as in \autoref{fig:mean_main}. Errorbars shown give the standard error of the dispersion. }
  \label{fig:std_main}
\end{figure}

\subsection{Alternate Models}
\label{subsec:alt_models}

To account for the current uncertainties in the galaxy scaling relations, in addition to the fiducial model we consider two alternative models.

In the first (``Alt Gas"), we vary the redshift dependence of the gas fraction by assuming a stronger evolution of the gas depletion time, $\tdep \propto (1+z)^{-0.6}$. This implies $n_z = 2.4$ for $z < 2$, and $n_z = 1.1$ for $z > 2$, in \autoref{eqn:fg}.

In the second (``Alt MMR"), we vary the redshift evolution of the MMR, adopting $\alpha_z = 0.6$ (see \autorefp{eq:mmr}), but keep the same gas fraction as in our fiducial model.

\subsection{Selecting Model Parameters}

We minimize the merit function $\cal M$ (\autoref{eqn:merit}) and find the best-fitting parameters $p_2$ and $p_3$ for each of the three models independently. The resulting best-fit parameters are presented in \autoref{tab:models}. In the next section we describe the results of the best variants of these models.

\section{Model predictions}
\label{sec:Results}

\cite{kravtsov_etal_2014} found that the stellar mass-halo mass relation of \cite{behroozi_etal_2013_main} overestimates the stellar mass of massive galaxies ($\Mstar \gtrsim 10^{12}\Msun$) due to over-subtraction of the intracluster light in the Sloan Digital Sky Survey from which the relation was derived. We therefore apply their corrections for $\Mstar(\Mh, z)$ at $z=0$, and use these corrected stellar masses for plotting the predicted present-day relations. Because the magnitude of the correction is unknown at higher redshift, when clusters actually form, we use the original relation of \cite{behroozi_etal_2013_main} for all calculations in the model. Throughout this section, we refer to $z=0$ halo and stellar masses as $\Mh$ and $\Mstar$ respectively. 

\subsection{Metallicity Trends: Mean and Dispersion}
\label{subsec:metallicity_trends}

\autoref{fig:mean_main} shows that the mean metallicity of observed GC systems increases 0.6~dex over the halo mass range $10^{11}\Msun$ to $\sim 10^{13}\Msun$. However, for $\Mh \gtrsim 10^{13}\Msun$, the mean $\feh$ no longer grows with increasing halo mass, but instead begins to decrease slightly ($\approx 0.1$~dex). This effect is caused by the fact that cluster formation peaks in halos with $\Mh \approx 10^{12}\Msun$ (see \autoref{subsec:formation_histories} for further discussion). Larger halos pass through this mass at higher redshift, and the redshift evolution of the MMR pushes the metallicity of any clusters that form then to lower values.

\autoref{fig:std_main} shows the width (standard deviation) of the metallicity distribution of GC systems over the same range of halo mass. Both observations and the model show a trend towards increasing GC metallicity dispersions in larger hosts. However, the normalization of the model is offset $\approx 0.1$~dex lower than the observed dispersions.

Both the fiducial and alternate models are very consistent in these trends.

\begin{figure}
\includegraphics[width=\columnwidth]{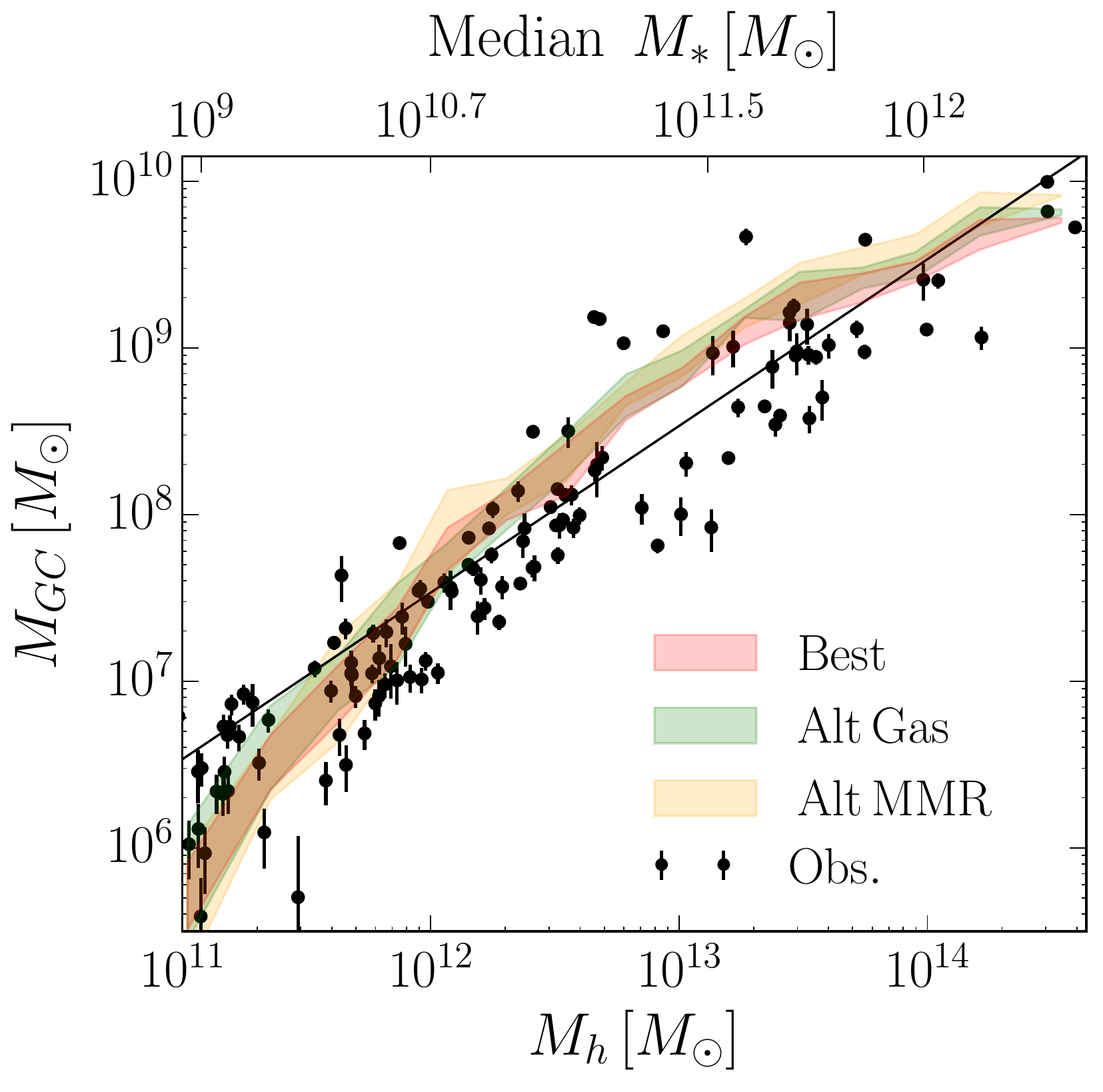}
\vspace{-5mm}
\caption{Observed and model trends for total GC system mass as a function of the host galaxy halo mass. Shaded regions give interquartile ranges of the models. Legend is as in \autoref{fig:mean_main}. The linear relation $\Mgc = 3.4\times 10^{-5}\Mh$ from \protect\citet{harris_etal_2015} is shown by the straight line.}
  \label{fig:m_main}
\end{figure}

\subsection{Globular Cluster System Mass - Halo Mass}
\label{subsec:mgc_mh}

\autoref{fig:m_main} shows that all three models consistently match the total globular cluster mass - halo mass relation. The average scatter around the linear trend in the model is 0.3~dex, in agreement with the 0.28~dex scatter in the observed relation \citep{harris_etal_2015, harris_etal_2017}. The observational fit includes also the 0.15-0.2~dex assumed uncertainty in halo mass, so that the intrinsic scatter could be smaller. In turn, our model has the galaxy-halo match uncertainty of 0.3~dex, which can completely account for the scatter in the $\Mgc-\Mh$ relation.

\begin{figure*} 
\includegraphics[width=0.9\textwidth]{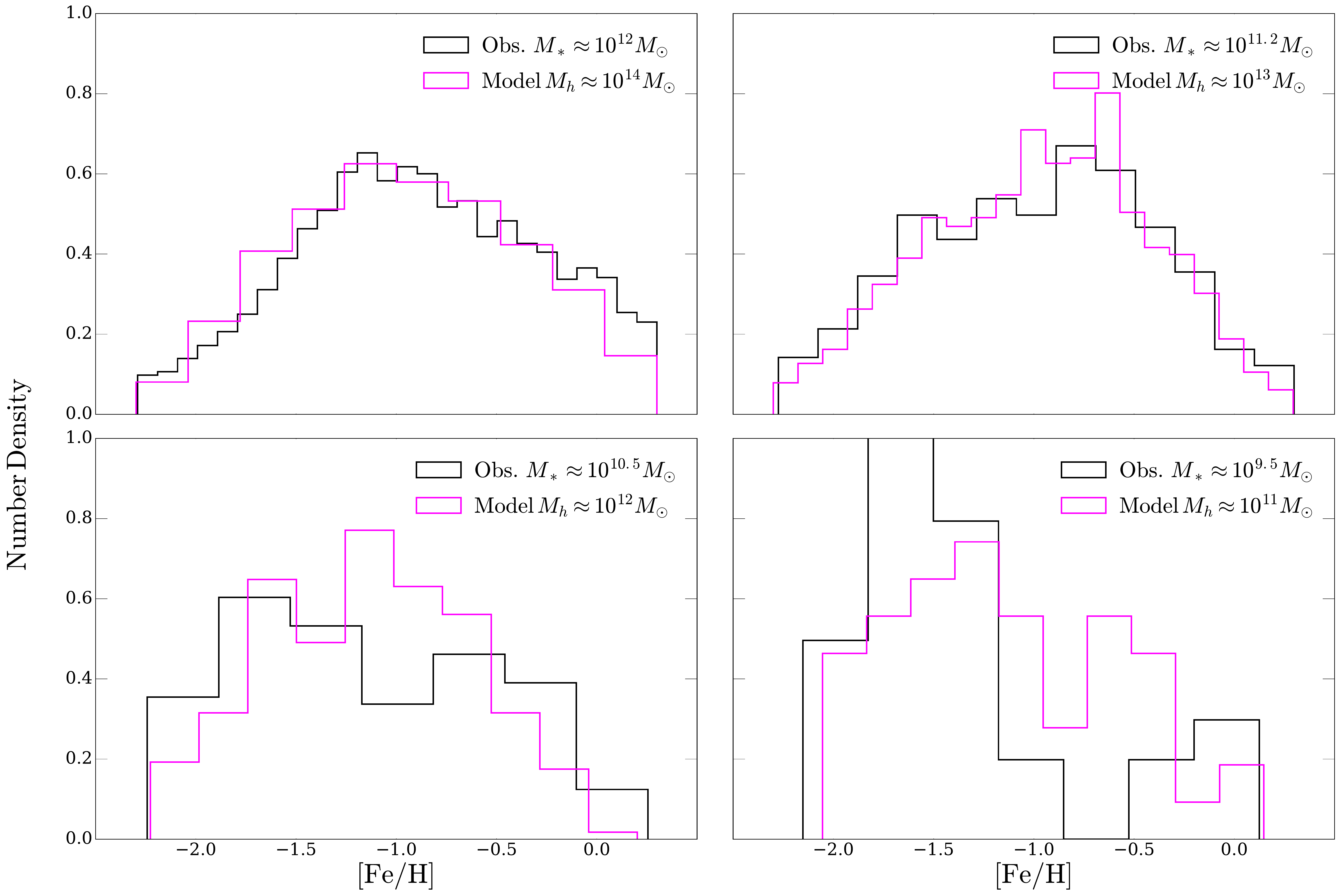}
\vspace{-2mm}
\caption{Normalized metallicity distributions for typical observed and model galaxies, varying by a factor of 10 in halo mass. Black represents observations, with the stellar mass of the galaxy indicated in the legend; magenta represents models, with the indicated halo mass. The KS probability values for each pair are 1\%, 50\%, 5\%, 3\%, in order of decreasing mass.}
  \label{fig:hist}
\end{figure*}

However, the models reveal a trend beyond linearity. There is \rev{noticeable curvature at low masses}. The observed points also follow this trend: $\Mgc$ clearly drops below the best linear fit (\autorefp{eq:mgcobs}) at $\Mh < 10^{12}\Msun$. \rev{Specifically, we find that the RMS error of the median model trend is a factor of $\approx 2$ lower than the RMS error of the best linear-fit at $M \lesssim 10^{12} \Msun$.} The model curves help to recognize this trend. 

The average scatter around the true mean trend in the model, rather than the linear relation, is even lower: 0.2~dex. Given the \rev{scatter in the stellar mass-halo mass relation}, we can assert that the derived non-linear $\Mgc-\Mh$ relation does not require any additional intrinsic scatter. 

The normalization of this relation is also remarkably close to the relation derived in the cosmological simulations of \cite{kravtsov_gnedin05}, which predicted $\Mgc/\Mh = (3-5)\times 10^{-5}$ for the halo mass range $10^9 - 10^{11}\Msun$ at redshift $z\approx 3$. However, this agreement may be coincidental, as it does not include the evolution of both $\Mgc$ and $\Mh$ until $z=0$. In our current model we find that the shape and scatter of the relation at $z=3$ are similar to the local relation, but with $\approx1$~dex higher normalization, i.e., $\Mgc\sim 3\times 10^{-4}\Mh$. At $z<3$, the GC system mass does not change appreciably, while the halo mass grows by an order of magnitude, thus causing the relation to shift rightward in the $\Mgc-\Mh$ plane.

\subsection{Globular Cluster Mass Function}
\label{sec:gcmf}

\revv{The cluster mass function in our model begins as a simple power-law $dN/dM \propto M^{-2}$. After $\approx$10~Gyr of disruption the shape of the mass function evolves into an approximately log-normal distribution by $z=0$. The median mass of the $z=0$ GC mass functions is very stable over all halos with $\Mh \gtrsim 10^{12} \Msun$ and located at $\log_{10}(M/\Msun) \approx 5.3$ for clusters with masses above our adopted completeness limit. This value is identical to the typical values inferred from observations of GC luminosity functions \citep[e.g.,][]{jordan_etal_2007}. At lower halo masses, $\Mh \lesssim 10^{12} \Msun$, the typical median mass of the $z=0$ mass functions is lower, $\log_{10}(M/\Msun) \approx 5.1$, with large scatter. Furthermore, the standard deviation of the mass functions scales with the mass of the host, ranging from 0.2~dex at $\Mh \approx 10^{11} \Msun$ to 0.6~dex for $\Mh \gtrsim 10^{13} \Msun$. \cite{jordan_etal_2007} found a similar range and scaling of the mass function width in the VCS galaxies (see their Fig.~14). The overall shape of the present-day mass functions is also broadly consistent with those of the observed systems in the VCS, with a robust goodness statistic of $G_M\approx 66\%$ across all three models. In a follow-up work we will present a detailed analysis of the adopted disruption rates and resulting cluster mass functions (N. Choksi \& O. Gnedin, in prep).}

\subsection{Multimodality of Globular Cluster Metallicity}
\label{subsec:multimodality}

\autoref{fig:hist} shows the comparison of the model predictions with the observed metallicity distribution in four galaxies of different mass at $z=0$. Less massive galaxies show clear bimodality (e.g., lower two panels of \autoref{fig:hist}). As $\Mh$ increases, the MDF converges towards the appearance of a broad, unimodal distribution.

To quantify these trends and split the distribution into red and blue subpopulations, we apply a Gaussian Mixture Modeling (GMM) routine developed in MG10 and model the metallicity distribution of each GC system as the sum of two gaussians. We take the cutoff between blue and red clusters to be the metallicity at which the fractional contributions of the two gaussians are equal. The median transition metallicity in our model systems occurs at $\feh = -0.88$, with a standard deviation of 0.18~dex among all halos. The median transition amongst the observed systems is lower, at $\feh = -1.1$, with a standard deviation of 0.25~dex. For systems in which the GMM routine returns a metallicity peak of the red subpopulation at $\feh < -1$, we instead take only the blue-peak. Such systems occur only in the smallest model halos, $\Mh \approx 10^{11} \Msun$, with only a few metal-poor clusters and no metal-rich clusters. 

\begin{figure}
\includegraphics[width=\columnwidth]{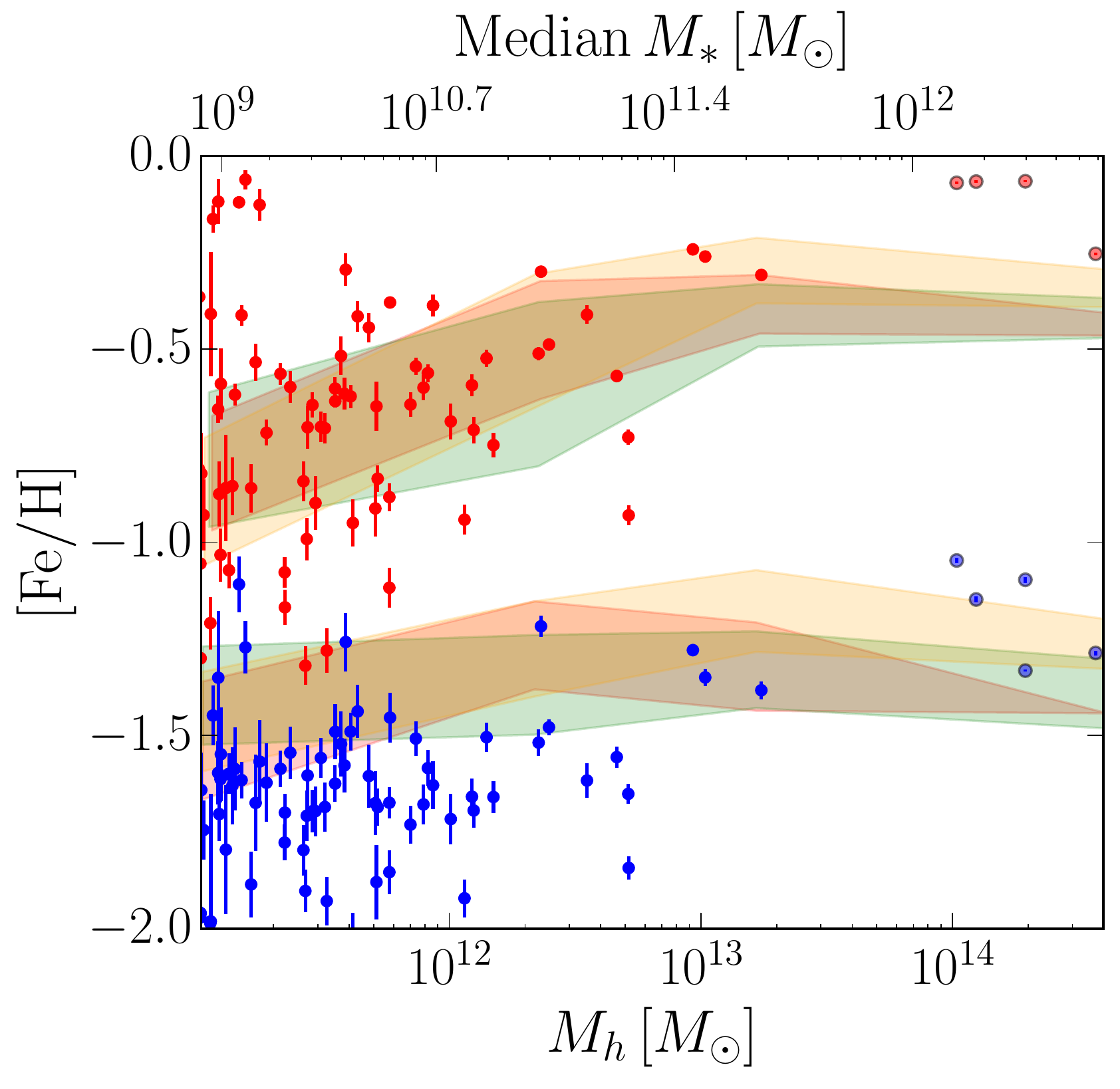}
\vspace{-5mm}
\caption{Location of peak metallicities of red and blue cluster subpopulations as a function of host halo mass. Shaded regions give the interquartile range of peak locations for the three models. Symbols with errorbars represent observed systems. Legend as in \autoref{fig:mean_main}.}
  \label{fig:peaks}
\end{figure}

\autoref{fig:peaks} shows that the peak metallicities of the red and blue clusters are very stable, varying only slightly with host halo mass. In agreement with observations of VCS \citep[e.g.,][]{peng_etal06}, we find the peak of the blue mode is essentially constant, while the peak of the red mode scales weakly with host mass. The best-fit linear relations with galaxy stellar mass are:
\begin{align}
  \feh_{\mathrm{red}} = (-0.48\pm 0.01) + (0.14\pm 0.02)\log_{10}\frac{\Mstar}{10^{10}\Msun} \nonumber\\ 
  \feh_{\mathrm{blue}} = (-1.35\pm 0.01) + (0.03\pm 0.01)\log_{10}\frac{\Mstar}{10^{10}\Msun}
 \label{eqn:model_peaks}
\end{align}
with intrinsic scatter $\sigma_{\mathrm{red}} = 0.18\pm 0.01$~dex and $\sigma_{\mathrm{blue}} = 0.16\pm 0.01$~dex.

For comparison, the observed relations for host galaxies with $\Mstar > 10^9 \Msun$ are:
\begin{align}
  \feh_{\mathrm{red}} = (-0.62\pm 0.02) + (0.12\pm 0.03)\log_{10}\frac{\Mstar}{10^{10}\Msun} \nonumber\\ 
    \feh_{\mathrm{blue}} = (-1.58\pm 0.02) + (0.08\pm 0.02)\log_{10}\frac{\Mstar}{10^{10}\Msun}
    \label{eqn:obs_peaks}
\end{align}
with intrinsic scatter $\sigma_{\mathrm{red}} = 0.28\pm 0.02$~dex and $\sigma_{\mathrm{blue}} = 0.19\pm 0.01$~dex. 

\begin{figure}
\includegraphics[width=\columnwidth]{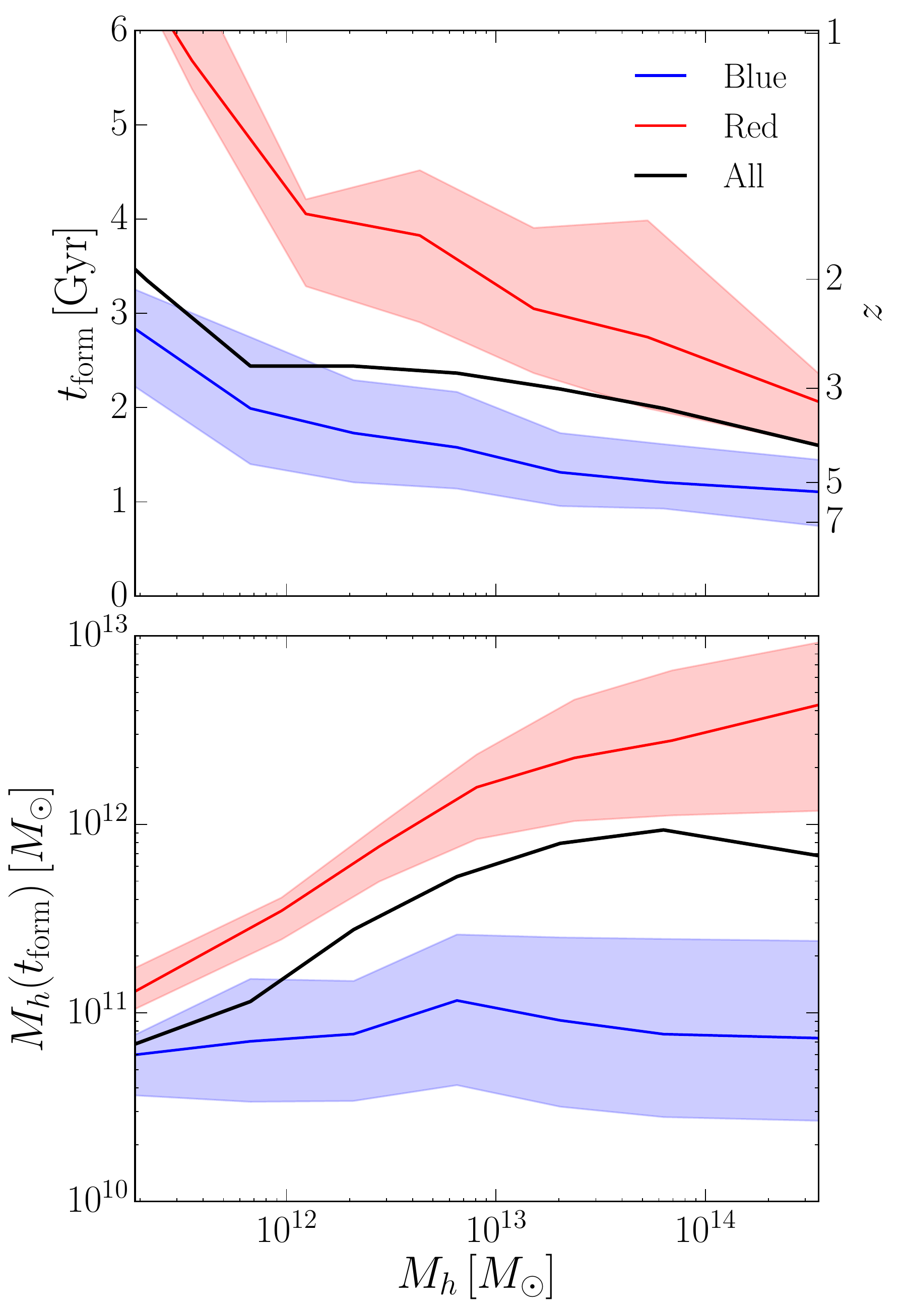}
\vspace{-5mm}
\caption{The redshift and cosmic time of cluster formation (\textit{top panel}), and host halo mass at time of formation (\textit{bottom panel}), in bins of halo mass at $z=0$. Specifically, we compute the interquartile range for the GC system in each halo, and plot the median range for all halos in the bin. Red and blue clusters are shown separately. Solid lines give median trends. Black solid line shows the median trend of all clusters.}
  \label{fig:formation1}
\end{figure}

\subsection{Formation Histories}
\label{subsec:formation_histories}

To understand the origin of the above relations, we investigate the timing and environment of cluster formation. The only two variables that control metallicity assigned to clusters in our model are the host galaxy mass and the formation epoch, via the redshift-dependent MMR (\autorefp{eq:mmr}).

We split the halos in bins of mass at $z=0$, and calculate the distribution of cosmic times $\tform$ of the formation of all clusters in each bin's halos. We also record the mass of the host halo at cluster formation, $\Mh(\tform)$.

The upper panel of \autoref{fig:formation1} shows that the median redshift of cluster formation increases with galaxy mass, from $z\approx 2$ for $\Mh \approx 10^{11}\Msun$ to $z\approx 4$ for $\Mh \approx 3\times 10^{14}\Msun$. This redshift range corresponds to cosmic time from 1.5 Gyr to 3.3 Gyr. The later formation times in less massive systems is similar to the phenomenon known as ``downsizing" in galaxy formation.

There are also clear systematic differences in the typical formation epochs between red and blue GCs. The bulk of blue clusters form between $z\approx 6$ and $z\approx 3$. Red clusters form over a more extended period, from $z\approx 4$ to $z\approx 0.8$. The typical offset between the formation of blue and red clusters is between 2 and 4~Gyr. 

The shaded regions in \autoref{fig:formation1} show only the 25-75\% percentiles of the distributions, and some blue clusters form later than the top blue contour, while some red clusters form earlier than the bottom red contour. We find that about 20\% of red and blue clusters form concurrently. Specifically, we find an overlap time such that the same fraction of blue clusters form after it as the fraction of red clusters that form before it. The overlap time ranges from $\tform = 1.6$~Gyr to 4.7 Gyr, increasing with decreasing halo mass. The trend with halo mass is similar to the median line shown in upper panel of \autoref{fig:formation1}. The overlap fraction of clusters, on the other hand, is consistent across the halo mass range and stays in the range between 15\% and 23\% .

The bottom panel of \autoref{fig:formation1} shows that the host halo mass in which GCs tend to form also differs strongly between red and blue clusters. Hosts of blue clusters are confined to a narrow range $3\times 10^{10} - 2\times 10^{11}\Msun$. In contrast, the host halo mass of red clusters increases significantly with final halo mass, and dominate the median trend for all clusters.

The model prediction can be compared to the results of \citet{behroozi_etal_2013_main} showing that overall star formation rates in galaxies peak near a halo mass $\approx 10^{12}\Msun$ (this trend is explicitly embedded in our model). Our model shows that the peak mass for forming GCs is somewhat lower (by a factor $2-3$) than that for all stars. Given that the bulk of stars in galaxies have higher metallicities than GCs, it is expected that GCs would form in lower-mass halos. The magnitude of the offset is relatively small, which indicates that GC formation is distinct but not completely different from overall star formation in galaxies.

\begin{figure}
\includegraphics[width=\columnwidth]{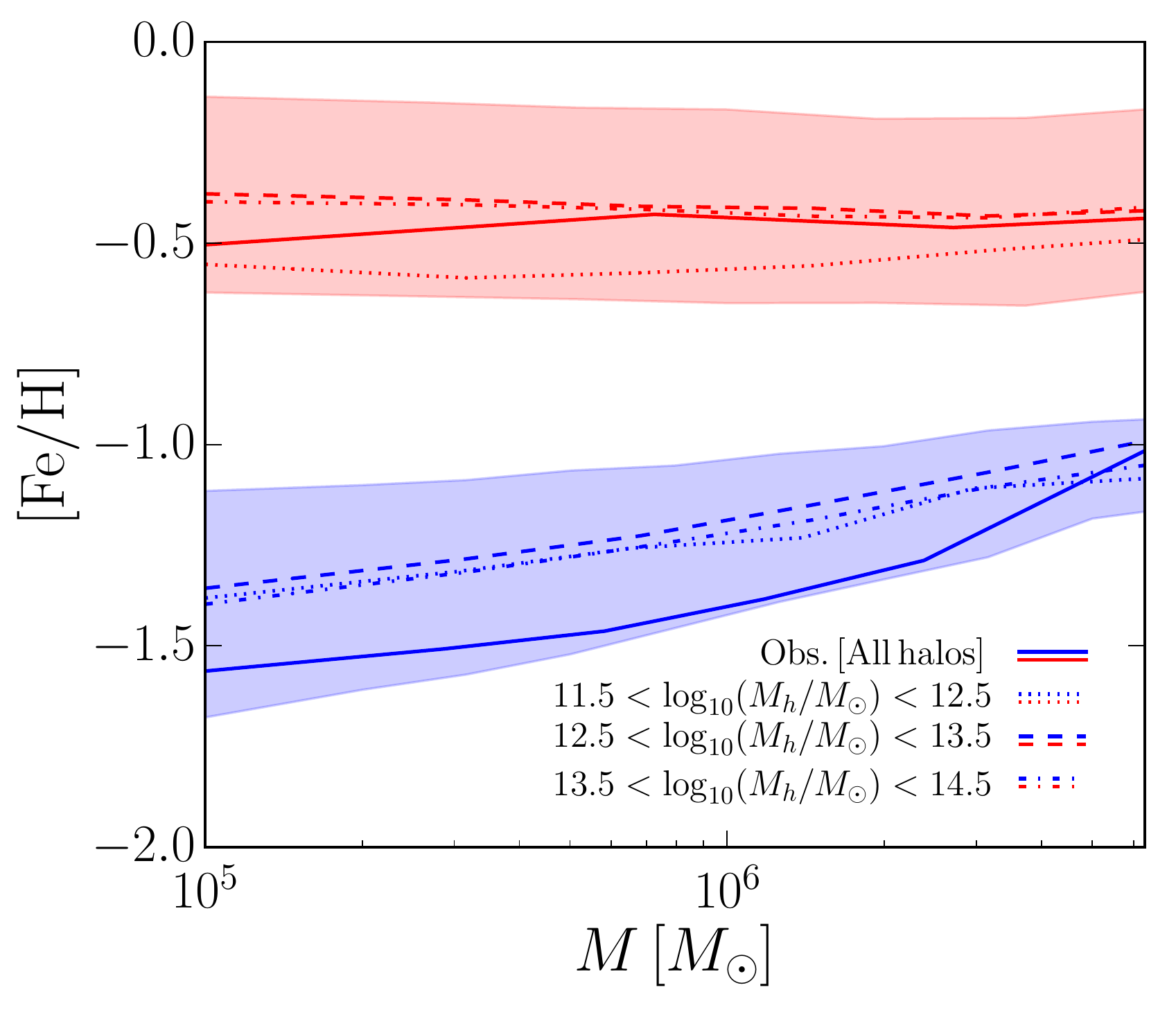}
\vspace{-5mm}
\caption{Metallicity of individual GCs as a function of cluster mass, separately for red and blue clusters. Shaded regions give the interquartile range for clusters stacked from  all the halos, while dotted, dashed, and dot-dash lines correspond to the mean trends for three bins of halo mass. For comparison, we include the observed trends from the Virgo Cluster Survey (solid lines).}
  \label{fig:blue_tilt_all}
\end{figure}

\begin{figure}
\includegraphics[width=\columnwidth]{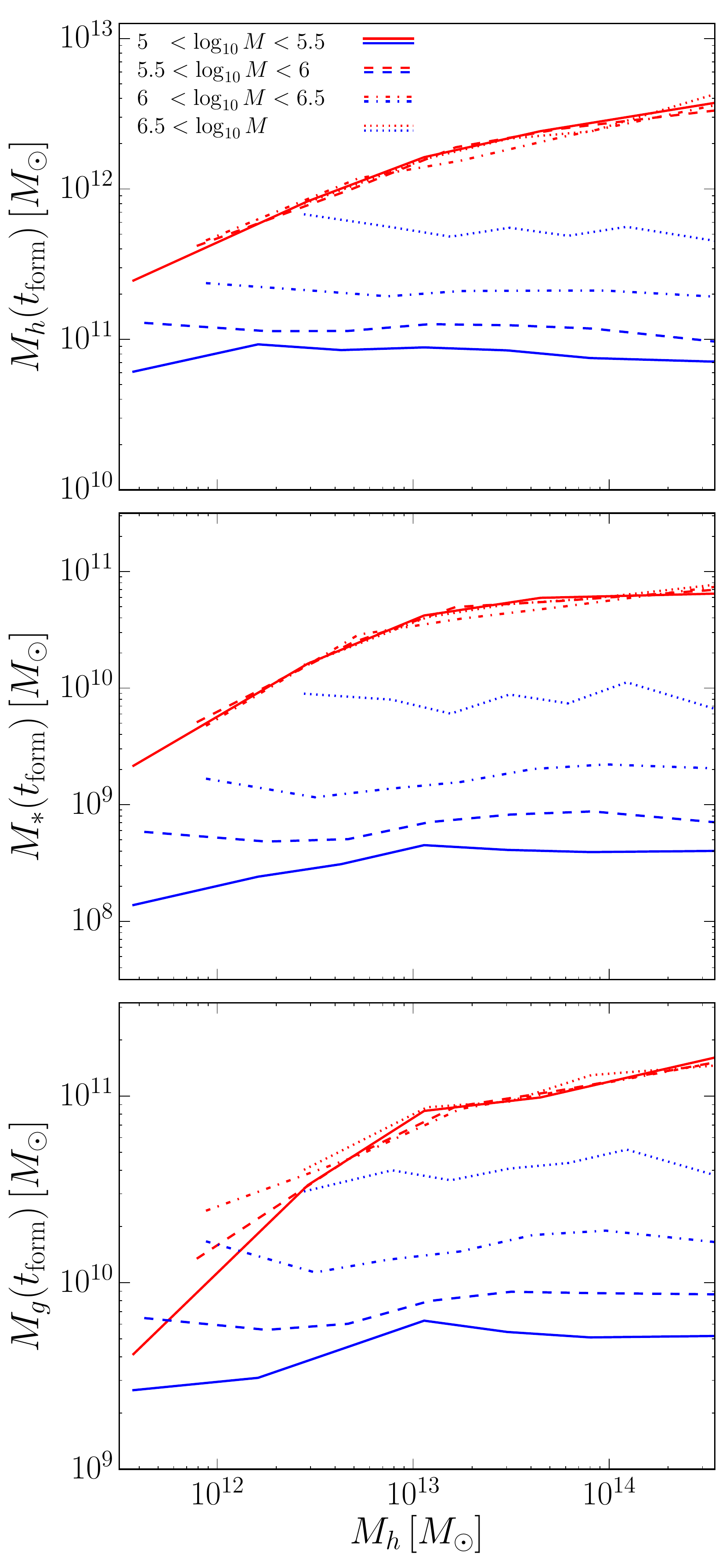}
\vspace{-5mm}
\caption{Median halo, stellar, and gas masses of cluster hosts at the time the cluster formed, split in four bins of individual cluster mass, separately for red and blue clusters. Combined, these explain the blue tilt in massive clusters (see \autoref{subsec:blue_tilt} for further discussion).}
  \label{fig:formation2}
\end{figure}

\subsection{Blue Tilt of Massive Clusters}
\label{subsec:blue_tilt}

Observations over the past decade revealed a subtle but statistically significant scaling of the metallicity of blue clusters with cluster mass: $Z \propto M^{\alpha}$. It has been called the ``blue tilt", in reference to the apparent tilt when cluster color is plotted against its magnitude. The color is proxy for metallicity, and the magnitude is proxy for mass. The slope $\alpha$ varies significantly between galaxies, ranging from 0.2 to 0.6, with some galaxies showing no trend at all \citep[][]{harris_etal06,  strader_etal06,  spitler2006, schiavon_etal_2013, fensch_etal_2014,harris_etal17_bcg3}. Only blue clusters more massive than $\sim 10^6\Msun$ show this tilt; for lower-mass clusters, which constitute the vast majority, there is no systematic trend. There is also no discernible correlation for red clusters. Because the high-mass clusters involved are rare, the blue tilt can only be detected in large enough samples of giant galaxies.

The first explanation for the blue tilt was suggested by \citet{strader_smith_2008} and \citet{bailin_harris09}, as being due to self-enrichment by supernova ejecta material in the course of cluster formation. Such an internal process seemed reasonable, because only massive clusters show the trend. Since other clusters in the same environment do not have it, it should not depend on global properties of the host galaxy. 

Our simple model does not include self-enrichment. Nevertheless, we found that it does predict the blue tilt effect!

\autoref{fig:blue_tilt_all} shows the result of stacking all our model samples. There is a clear correlation for blue clusters. A linear relation for the mean $\feh$ can be fit as:
\begin{equation}
  \feh_{\mathrm{blue}} = (-1.24 \pm 0.02) + (0.23 \pm 0.01)\log_{10}\frac{M}{10^6\Msun}
\end{equation}
for $M > 5\times 10^{5}\Msun$, with intrinsic scatter $\sigma = 0.27\pm 0.01$~dex. Remarkably, there is almost no dependence on halo mass. The dashed, dotted, and dash-dotted lines show the sample split into bins of host halo mass -- they follow essentially the same relation. 

In comparison, the observed trend for the VCS GCs is:
\begin{equation}
   \feh_{\mathrm{blue}} = (-1.41 \pm 0.01) + (0.30 \pm 0.07)\log_{10}\frac{M}{10^6\Msun}
\end{equation} 
in the same mass range, with intrinsic scatter $\sigma = 0.33\pm 0.01$~dex. The model result is therefore within ``one-sigma" of the observed scaling. For red clusters there is no statistically significant trend in either model or observations.

For individual halos, the model produces a distribution of mass-metallicity relations for the blue clusters. The range of slopes, $\alpha$, extends between $\alpha=0$ (i.e., no blue-tilt) and $\alpha \approx 0.55$, with a peak near $\alpha \approx 0.3$.  

\autoref{fig:formation2} illustrates the origin of the blue tilt in our model. Halos with larger cold gas reservoirs can generate more massive clusters, because they more fully sample the cluster initial mass function (\autorefp{eq:cimf}). These halos will also have higher stellar mass, so the clusters they form inherit a higher metallicity. 

Specifically, the top panel of \autoref{fig:formation2} shows that the median halo mass at the time of blue cluster formation, $\Mh(\tform)$, is independent of the final halo mass but is strongly increasing with individual cluster mass. Clusters with $M > 10^{6.5}\Msun$ simply do not form in halos less massive than $4\times 10^{11}\Msun$. Similarly, the contours of gas mass $M_g(\tform)$ in the bottom panel are essentially flat but shifted vertically.  The smallest clusters have median hosts with $M_g(\tform) \sim (2-4)\times 10^{9}\Msun$, which according to our model (\autorefp{eqn:mgc} with $p_2 = 6.75$) create a GC population with combined mass $\Mgc \sim 3\times 10^{6}\Msun$. Given that clusters lose about half of their mass due to stellar evolution and some more due to tidal disruption, only clusters with $M \lesssim 10^{6}\Msun$ could be produced in such hosts. Clusters with $M > 10^{6}\Msun$ require $M_g(\tform) > 10^{10}\Msun$, as is demonstrated by the dotted and dot-dashed lines. These massive hosts also have larger stellar masses, $\Mstar(\tform)$, shown in the middle panel. Through the galactic MMR, they imply higher metallicity for higher mass clusters. Thus the emergence of blue tilt in our model is not due to an internal process (such as self-enrichment) but instead is due to lack of massive, low-metallicity clusters, which cannot be produced by low-mass metal-poor hosts.

There is no such trend for red clusters because they form at later times in more massive halos. Such halos have sufficient amounts of gas available for cluster formation, so the cluster initial mass function is fully sampled and more massive GCs are equally likely to form in all environments.

\begin{figure}
\includegraphics[width=\columnwidth]{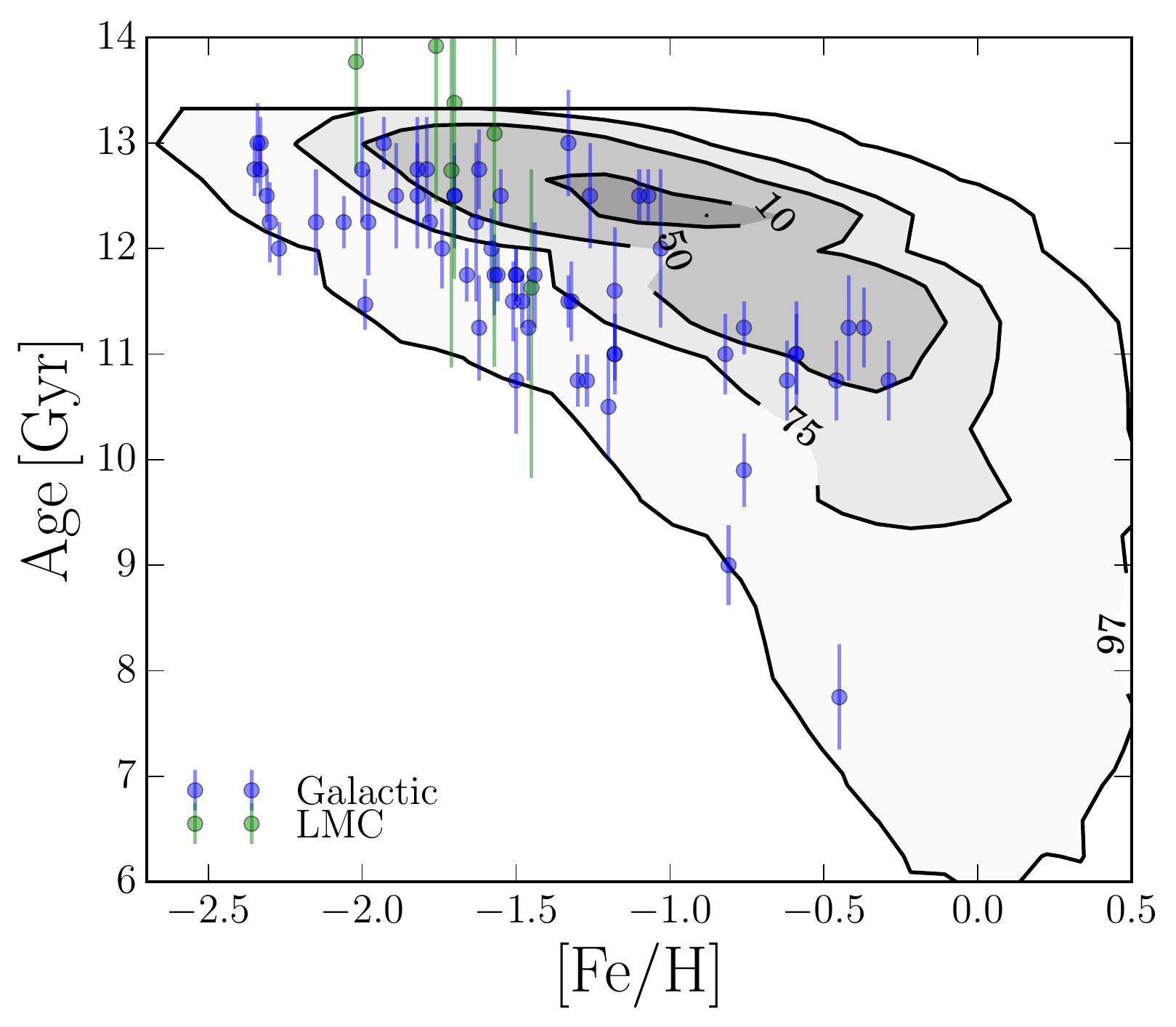}
\vspace{-5mm}
\caption{Age-metallicity distribution of all model clusters in the fiducial model. Contours show the parameter space occupied by a given percentage of model clusters, shown as numbers on the contour lines. For comparison, we show the observed ages for the Galactic globular clusters \citep[blue points with error-bars, from][]{leaman_etal13} and six LMC clusters \citep[green points, from][]{wagner-kaiser_etal_2017}.}
  \label{fig:age_main}
\end{figure}

\subsection{Age-Metallicity Relation}
\label{subsec:age_met}

The formation of clusters with higher metallicity in more massive host galaxies naturally results in an age-metallicity relation (AMR) of GCs. Such an AMR was predicted by all earlier versions of our model, in MG10 and LG14 (see Fig. 15 of LG14). Similarly, we find it in the new model with little change.

\autoref{fig:age_main} shows a systematic trend for clusters with $\feh > -0.5$ to be younger by $2-4$~Gyr than the metal-poor ones. The only difference with the LG14 result is minor, in that the innermost contour of 25\% oldest clusters is narrower near $12-13$~Gyr. Thus, the cluster AMR is a robust prediction of our model.

The model prediction is consistent with the best available age measurements based on resolved stellar populations for Galactic GCs \citep{dotter_etal11, vandenberg_etal13, leaman_etal13} and LMC clusters \citep{wagner-kaiser_etal_2017}, as well as integrated light spectra for GCs in the nearby elliptical galaxies (not shown) by \citet{georgiev_etal12}. The tail of higher-metallicity clusters $\feh > -0.5$ is expected in the systems of giant elliptical galaxies. Our model makes a clear prediction that those clusters should be systematically younger than their metal-poor cousins.

\subsection{Cluster Fraction at Different Metallicities}
\label{subsec:sn}

Observations of stellar halos in M31 and the elliptical galaxy Cen A show that the fraction of halo stars in a given metallicity range that are contained in GCs increases with decreasing metallicity \citep[e.g.,][]{lamers_etal_2017}. This fraction also does not depend on the distance from the galaxy center. While our model has no spatial information of clusters, it calculates the cluster and galaxy metallicity. Therefore, we can investigate whether the model reproduces the overall trend.

We first compute the total mass of clusters in a given metallicity bin, $M_{\mathrm{GC,bin}}$. To compute the mass of field stars in the bin, we track the stellar masses and metallicities of the main progenitor branch (MPB; defined as the track along the merger tree with the largest integrated mass history) of each $z=0$ halo. We then record the stellar mass at each output where the MPB stellar mass crosses the edge of a given metallicity bin. The stellar mass in a given  bin with lower and upper edges $\feh_i$ and $\feh_{i+1}$ is:
\begin{equation}
M_{\star,\mathrm{bin}} = \Mstar(\mathrm{[Fe/H]}_{i+1}) - \Mstar(\mathrm{[Fe/H]}_{i}).
\end{equation}
The mass of field stars is the full stellar mass minus the mass of GCs: $M_{\mathrm{field,bin}} = M_{\star,\mathrm{bin}} - M_{\mathrm{GC,bin}}$.

\begin{figure}
\includegraphics[width=\columnwidth]{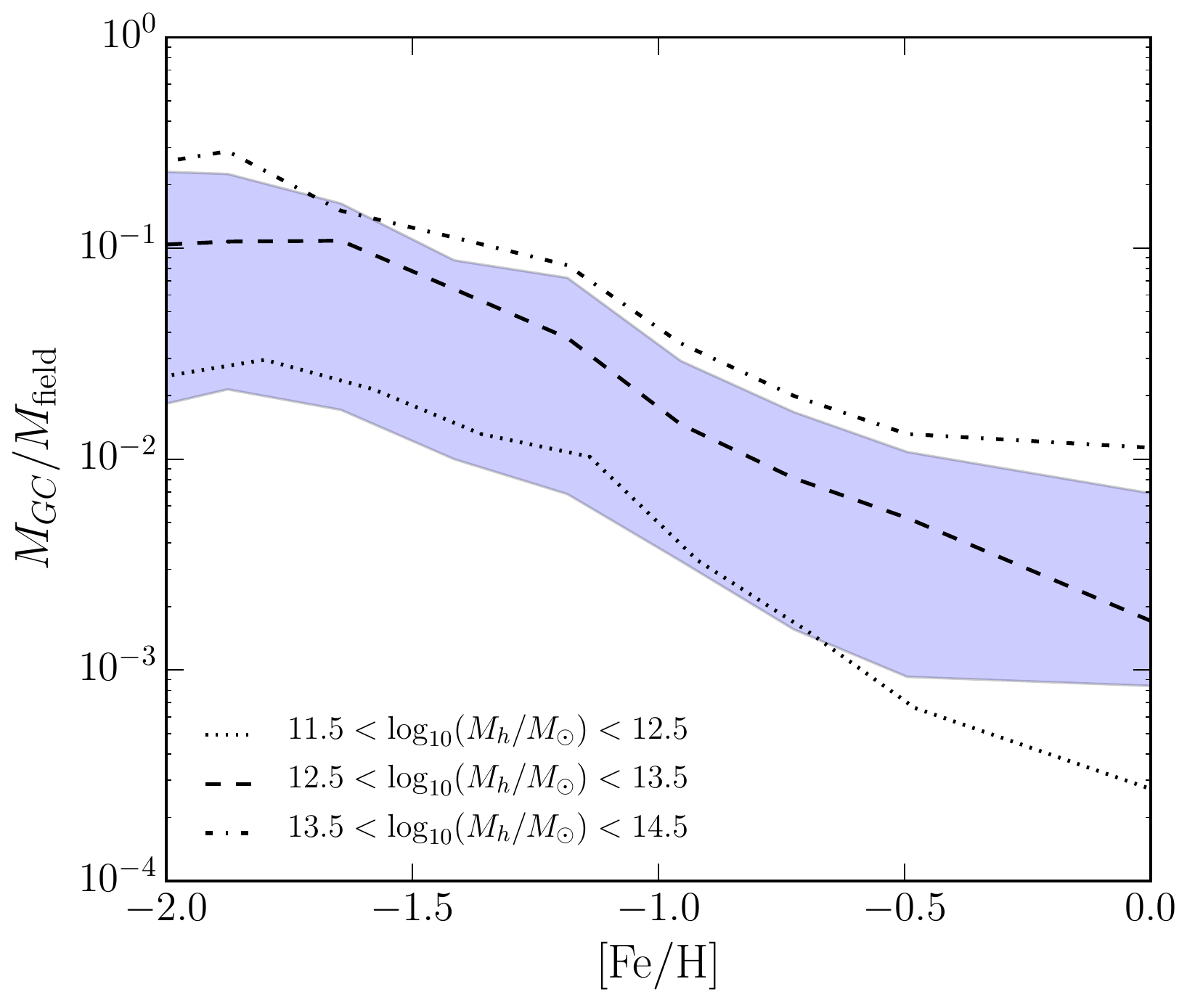}
\vspace{-5mm}
\caption{Ratio of mass in clusters to mass in field stars in different metallicity bins. The shaded region shows the interquartile for all model halos, and lines show the trend for three bins of halo mass.}
  \label{fig:sn}
\end{figure}

In \autoref{fig:sn} we show the result of this process. As in observations, the cluster fraction increases with decreasing metallicity. We cannot directly plot the observed ratios because the observations only cover a small subset of halo stars, and therefore do not give the correct normalization. However, the direction and magnitude of the predicted trend are comparable to the observations. The cluster fraction varies by about two dex over the range $-2 < \feh < 0$. Separation into bins of halo mass shows that the general trend is present over all host masses. At the same time, more massive halos have a systematically higher normalization of their cluster fraction.

\section{Discussion}
\label{sec:discussion}

We have presented an updated model of globular cluster formation and evolution in galaxies of a vast mass range, from dwarfs to giants. It matches the observed number and mean metallicity of GC systems, and predicts the ``blue tilt" mass-metallicity relation for massive blue clusters. Yet, one of the clear differences that remains between the best model and observations is the normalization of the metallicity dispersion $\sigma_Z$ (\autoref{fig:std_main}). The offset is $\approx 0.1-0.2$~dex. None of the alternative models that we tried is able to reduce this discrepancy. 

We find that decreasing $p_3$ to 0.2 can match the normalization of $\sigma_Z$, because it allows for more cluster formation events. However, this leads to increased cluster formation at later times, which increases the mean cluster metallicity above the observations. Boosting $p_2$  can increase cluster formation at \textit{earlier} times, and therefore lower the mean metallicity, but this then leads to a normalization of the $\Mgc-\Mh$ relation that is above the observed relation by $\approx 0.5$ dex. It was not possible to match the normalization of $\sigma_Z$ and produce reasonable results in matching the other observational constraints.

Despite the different dispersions, the mean metallicity of GC systems in halo mass bins is consistent between the model and observations, except for the highest mass bin. In that bin the mean metallicity decreases with halo mass, because of earlier formation of clusters. The observed mean declines even more than the model, by an additional 0.15 dex, and has the $t$-test probability of being drawn from the same distribution $p\sim 10^{-3}$. All other bins are consistent. 

\subsection{Numerical Resolution and Formation Epoch of First Clusters}

The upturn in $\tform$ for low-mass ($\sim 10^{11}\Msun$) halos, shown in \autoref{fig:formation1}, could in principle be due to the numerical resolution of our adopted simulation. If these low-mass halos contained insufficient number of particles to be identified by the halo finder algorithm at higher redshift, cluster formation would be pushed to later times.
 
The \textit{Illustris} catalogs include halos as small as $\Mh > 10^{8}\Msun$. These halos are resolved with 13 particles. However, we find that the minimum halo mass in which \textit{surviving} clusters form in our model is $\Mh \approx 10^{9.5} \Msun$, which are resolved with $\approx$400 particles. Therefore, these halos should be present in the merger trees of $\Mh(z=0) \lesssim 10^{11.5} \Msun$ halos if they truly existed, and therefore the downsizing effect in Fig. \ref{fig:formation1} is likely not due to numerical resolution.

\subsection{Uncertainty in Scaling Relations and Alternative Models}
\label{sec:alt_discussion}

In the Alt-MMR model, basic trends remain the same as in the fiducial model shown in \autoref{fig:formation1}. To compensate for the weaker redshift evolution of the MMR, cluster formation is pushed to higher redshift and lower-mass halos: the average $\tform$ decreases by 0.4~Gyr, while the average $\Mh(\tform)$ decreases by a factor of 2. The ``downsizing" effect is still present, but is weaker than in the fiducial model. Few red clusters form after $z \approx 2$, whereas in the fiducial model red clusters form until $z \approx 1$.

The Alt-Gas model also shows little change in the average $\tform$ and $\Mh(\tform)$. This model differs from the fiducial model only in lower gas fractions at high redshift. So, most differences are eliminated by an appropriate boost to $p_2$ (at the same value of $p_3$) which increases the gas content available for cluster formation over all redshifts (via \autorefp{eqn:mgc}). 

Note that the simple functional form of the gas fraction $\eta(\Mstar,z)$ that we adopted in \autoref{eqn:fg} necessarily ignores more subtle variations. The gas fraction presented in Figure~4 of \citet{tacconi_etal_2017} falls below the single power-law fit at $\Mstar < 2\times 10^{9}\Msun$, which could lead to fewer blue clusters forming in low-mass galaxies. This would push the formation of blue clusters to even later epochs. Also, the galaxy-to-galaxy scatter of the gas fraction at fixed mass could spread cluster formation over a longer period. The model predictions could be improved when the scaling relations include a larger dataset of galaxies with measured molecular gas content in the redshift range $z=2-5$, corresponding to the formation epoch of most globular clusters.

\subsection{Comparison with LG14 Model}

Despite the many updates to the galaxy scaling relations relative to the LG14 model, our new results are broadly consistent with the old ones. Switching to the \textit{Illustris} halo catalogs allowed us to use more frequent outputs, improve the merger criterion, and apply it to a wider variety of galaxies. 

A more substantial difference is changing the disruption rate calculation in the tidally-limited regime, from $\ttid \propto M$ to $\ttid \propto M^{2/3}$, and adding the disruption in weak tides regime. That has changed the resulting cluster mass function at $z=0$, but correcting the disruption rate with the factor $P = 0.5$ brought the mass distribution back in agreement with the data. Stronger disruption of high-mass clusters then required a larger value of $p_2$ in the best model.

The only other noticeable change is the extension of the age-metallicity relation for the largest galaxy systems, where the oldest clusters are now present at higher metallicity, $\feh \approx -0.5$. The general trend of the age-metallicity relation is very robust and consistent with all previous versions of the model.

\subsection{Comparison with Other Models}

\citet{kruijssen12, kruijssen2015} proposed an analytical model for the formation of GCs in high-pressure regions of gas-rich galaxies, similar to the formation of young star clusters in the nearby universe. Many aspects of that model are analogous to ours. Clusters form within high-redshift disk galaxies, but at a fixed epoch taken to be $z=3$, and then migrate to the galactic halo after major mergers. Cluster metallicity is assigned from the galactic MMR at the formation epoch.

The role of mergers in that model is somewhat different from our model. They not only help create clusters by increasing the galactic ISM pressure, but also destroy some young clusters by time-variable tidal perturbations (``shocks") in the dense ISM. Other clusters are gravitationally scattered into the galactic halo and survive for much longer times because of the weaker tidal field. Our model currently does not account for tidal shocks and therefore mergers only promote the formation of globular clusters.

The \citet{kruijssen12} model was recently incorporated on a sub-grid level in the hydrodynamic simulations of galaxy formation based on the EAGLE model. This new E-MOSAICS simulation \citep{pfeffer_etal_2017} produces ten realizations of disk-dominated galaxies with total mass $\Mh\sim 10^{12}\Msun$. The details of the calculation of final cluster properties in this model are different from our model. They include explicit calculation of the gas cooling, star formation, and stellar and AGN feedback in galaxies within dark matter halos. The formation of star clusters is implemented on the spatial scale smaller than the whole galaxy but larger than individual giant molecular clouds (GMCs), and is based on the local gas pressure and Toomre stability criterion. The disruption of clusters is also different from our calculation, and includes tidal truncation and tidal shocks based on the local gravitational field calculated of the scale $\sim 350$~pc. Yet, despite all these differences from our model, the E-MOSAICS model predicts similar timing ($z\approx 1-4$) for the formation of massive clusters capable of becoming globular clusters.

Both models share common predictions about the properties of globular cluster systems. The age-metallicity relation follows necessarily from the observed mass-metallicity relation of host galaxies and the assumption that globular clusters inherit the average metallicity of their hosts at the time of formation. Metal-rich clusters are, on average, several Gyr younger than the metal-poor ones. Future measurements of GC ages in external galaxies, accurate to $\la 1$~Gyr, would be most important for testing this robust model prediction.

The role of galactic mergers in GC formation has also become better understood. Gas-rich mergers promote the formation of massive clusters by compressing the ISM to higher density and pressure and creating GMCs. This conclusion is also supported by detailed hydrodynamic simulations of cluster formation by \citet{li_etal17, li_etal18}. In these ultrahigh-resolution cosmological simulations, star clusters form in GMCs within high-redshift ($z>1.5$) galaxies, and terminate their growth by the momentum, energy, and radiation feedback of their own stars. Clusters form continuously at all times, but galactic mergers enhance the formation of the largest GMCs and most massive clusters.

\subsection{Comparison with Observations of High-Redshift Galaxies}

\citet{shapiro_etal_2010} discussed the possibility of formation of metal-rich GCs in luminous clumpy galaxies detected in H$\alpha$ at $z\approx 2$ \citep{forster_schreiber_etal_2009, forster_schreiber_etal_2011}. These clumps represent star-forming complexes on the scale of about 1 kpc, whose luminosity may be dominated by a few massive star clusters. At higher resolution, either due to amplification by gravitational lensing \citep{johnson_etal_2017, vanzella_etal_2017} or using rest-frame FUV observations \citep{soto_etal_2017}, the clumps appear to fragment into smaller pieces. The metallicity and kinematics of these clumps are consistent with typical red globular clusters. 

The range of stellar mass $\Mstar \sim 10^{10}-10^{11}\Msun$ of red GC hosts in our model (see bottom panel of \autoref{fig:formation2}) is consistent with the observed masses of these $z\approx 2$ galaxies. The observed redshift range also matches the typical epoch of the formation of red clusters, shown in the top panel of \autoref{fig:formation1}. Thus these observed clumps could indeed represent young proto-globular clusters. 

\section{Conclusions} 
\label{sec:conclusions}

We have extended the model of globular cluster formation and evolution using dark matter halo merger trees from the \textit{Illustris} simulation and updated scaling relations for the cold gas fraction and its metallicity at high redshift. Our main results are summarized below.
\begin{enumerate}
\item The mean metallicity and dispersion of GC systems increases weakly with host halo mass. For hosts with $\Mh \gtrsim 10^{13}\Msun$, the mean metallicity flattens at $\feh\approx -0.9$ (\autoref{fig:mean_main}-\ref{fig:std_main}).

\item The GC system mass-halo mass relation is a robust prediction of the model. The model trend is non-linear, but matches the data even better than a linear relation (\autoref{fig:m_main}). The residual scatter decreases from 0.3~dex for the linear fit to 0.2~dex for the non-linear fit.

\item The wide range of observed GC metallicity distributions -- from bimodality in small systems to unimodality in the largest systems -- is reproduced by the model. The mean metallicity of the blue clusters is nearly constant with host halo mass, while that of the red clusters scales weakly with host mass (\autoref{fig:peaks}).

\item The model predicts distinct, but overlapping, formation times and locations of the blue and red clusters; the former form at $z\approx 5-7$ in small halos, while the latter form at $z\approx 2-4$ in more massive halos (\autoref{fig:formation1}).

\item As a result of (iv), a mass-metallicity relation naturally arises for the blue clusters (\autoref{fig:blue_tilt_all}). This trend occurs because their formation sites have much smaller total gas mass. Massive clusters can only form in halos that have large gas reservoirs; such halos will also have higher stellar masses and therefore higher metallicities (\autoref{fig:formation2}). 

\item The age-metallicity prediction of the model is very robust to changes in scaling relations (\autoref{fig:age_main}). Most metal-rich clusters are several Gyr younger than the metal-poor clusters.

\item The fraction of galaxy stellar mass locked in GCs is a strong function of metallicity and halo mass (\autoref{fig:sn}). The GC fraction can reach $\sim 10\%$ at $\feh < -1.5$.
\end{enumerate}

We provide an online table of our compilation of observational data used in Figs 1-3. The columns are: galaxy ID, galaxy stellar mass, mean $\feh$ of the GC system and its standard error, $\feh$ dispersion and its standard error, and locations of the red and blue peaks of the full GC metallicity distribution.

\section*{Acknowledgements}

We thank Bill Harris for numerous discussions of globular cluster systems in the HST-BCG survey and the color-metallicity calibration, Peter Behroozi for helpful conversations about the stellar mass-halo mass relation, and Goni Halevi for very helpful commentary throughout the course of this work. This work was supported in part by NASA through grant NNX12AG44G and by NSF through grant 1412144.




\bibliographystyle{mnras}
\bibliography{GC/gc_oleg,GC/gc_nick} 






\bsp	
\label{lastpage}
\end{document}